\documentclass[draftcls,onecolumn]{IEEEtran}
\usepackage{amsfonts}
\usepackage{amssymb}
\usepackage{amsmath,epsfig}
\usepackage{cite}
\usepackage{graphicx}
\usepackage{nccmath}
\usepackage{lscape}
\usepackage{graphics}
\usepackage{subfigure}
\usepackage{dsfont}

\input{tcilatex}
\graphicspath{{figures/}}

\begin{document}

\title{Some results on the Weiss-Weinstein bound for conditional and
unconditional signal models in array processing}
\author{ Dinh Thang VU, Alexandre RENAUX, R\'{e}my BOYER, Sylvie MARCOS 
\thanks{%
The authors are with Universit\'{e} Paris-Sud 11, CNRS Laboratoire des
Signaux et Syst\`{e}mes, Supelec, 3 rue Joliot Curie, 91192 Gif-sur-Yvette
Cedex, France (e-mail: \{Vu,Renaux,Remy.Boyer,Marcos\}@lss.supelec.fr)} 
\thanks{%
This project was funded by both r\'{e}gion \^{I}le de France and Digiteo
Research Park. Section \ref{Sec: Specific application PA Conditional model}
of this paper has been partially presented in \cite{VRBM10}.}}
\maketitle

\begin{abstract}
In this paper, the Weiss-Weinstein bound is analyzed in the context of
sources localization with a planar array of sensors. Both conditional and
unconditional source signal models are studied. First, some results are
given in the multiple sources context without specifying the structure of
the steering matrix and of the noise covariance matrix. Moreover, the case
of an uniform or Gaussian prior are analyzed. Second, these results are
applied to the particular case of a single source for two kinds of array
geometries: a non-uniform linear array (elevation only) and an arbitrary
planar (azimuth and elevation) array.
\end{abstract}

\begin{keywords}
Weiss-Weinstein bound, DOA estimation, array processing.
\end{keywords}

\section{Introduction}

Sources localization problem has been widely investigated in the literature
with many applications such as radar, sonar, medical imaging, etc. One of
the objective is to estimate the direction-of-arrival (DOA) of the sources
using an array of sensors.

In array processing, lower bounds on the mean square error are usually used
as a benchmark to evaluate the ultimate performance of an estimator. There
exist several lower bounds in the literature. Depending on the assumptions
about the parameters of interest, there are three main kinds of lower
bounds. When the parameters are assumed to be deterministic (unknown), the
main lower bounds on the (local) mean square error used are the well known
Cram\'{e}r-Rao bound \cite{Cra46} and the Barankin bound \cite{Bar49} (more
particularly their approximations \cite{MS69}\cite{MH71}\cite{Abe93}\cite%
{CGQL08}\cite{TT10a}). When the parameters are assumed to be random with a
known prior distribution, these lower bounds on the global mean square error
are called Bayesian bounds \cite{VTB07}. Some typical families of Bayesian
bounds are the Ziv-Zakai family \cite{ZZ69}\cite{BT74}\cite{BSEV97} and the
Weiss-Weinstein family \cite{ww85a}\cite{RO07}\cite{RFLRN08}\cite{TT10b}.
Finally, when the parameter vector is made from both deterministic and
random parameters, the so-called hybrid bounds have been developed \cite%
{RS87a}\cite{RM97}\cite{BGRBB08}\cite{NM09}.

Since the DOA estimation is a non-linear problem, the outliers effect can
appear and the estimators mean square error exhibits three distinct
behaviors depending on the number of snapshots and/or on the signal to noise
ratio(SNR) \cite{VT68}. At high SNR and/or for a high number of snapshots, 
\textit{i.e.}, in the asymptotic region, the outliers effect can be
neglected and the ultimate performance are described by the
(classical/Bayesian/hybrid) Cram\'{e}r-Rao bound. However, when the SNR
and/or the number of snapshots decrease, the outliers effect lead to a quick
increase of the mean square error: this is the so-called threshold effect.
In this region, the behavior of the lower bounds are not the same. Some
bounds, generally called global bounds (Barankin, Ziv-Zakai,
Weiss-Weinstein) can predict the threshold while the others, called local
bounds, like the Cram\'{e}r-Rao bound or the Bhattacharyya bound cannot.
Finally, at low SNR and/or at low number of snapshots,\textit{\ i.e.}, in
the no-information region, the deterministic bounds exceed the estimator
mean square error due to the fact that they do not take into account the
parameter support. On the contrary, the Bayesian bounds exploit the
parameter prior information leading to a "real" lower bound on the global
mean square error.

In this paper, we are interested in the Weiss-Weinstein bounds which is
known to be one of the tightest Bayesian bound with the bounds of the
Ziv-Zakai family. We will study the two main source models used in the
literature \cite{HLS93chap4}: the unconditional (or stochastic) model where
the source signals are assumed to be Gaussian and the conditional (or
deterministic) model where the source signals are assumed to be
deterministic. Surprisingly, in the context of array processing, while
closed-form expressions of the Ziv-Zakai bound (more precisizely its
extension by Bell et. al. \cite{BEV96a}) were proposed around 15 years ago
for the unconditional model, the results concerning the Weiss-Weinstein
bound are, most of the time, only conducted by way of computations.
Concerning the unconditional model, in \cite{NH88}, the Weiss-Weinstein
bound has been evaluated by way of computations and has been compared to the
mean square error of the MUSIC algorithm and classical Beamforming using a
particular $8\times 8$ element array antenna. In \cite{NVT94}, the authors
have introduced a numerical comparison between the Bayesian Cram\'{e}r-Rao
bound, the Ziv-Zakai bound and the Weiss-Weinstein bound for DOA estimation.
In \cite{Ath01}, numerical computations of the Weiss-Weinstein bound to
optimize sensor positions for non-uniform linear arrays have been presented.
Again in the unconditional model context, in \cite{XBR04}, by considering
the matched-field estimation problem, the authors have derived a semi
closed-form expression of a simplified version of the Weiss-Weinstein bound
for the DOA estimation. Indeed, the integration over the prior probability
density function was not performed. The conditional model (with known
waveforms) is studied only in \cite{Ren07}, where a closed-form expression
of the WWB is given in the simple case of spectral analysis and in \cite%
{VRBM10} which is a simplified version of the bound.

While the primary goal of this paper is to give closed-form expressions of
the Weiss-Weinstein bound for the DOA estimation of a single source with an
arbitrary planar array of sensors, under both conditional and unconditional
source signal models, we also provide partial closed-form expressions of the
bound which could be useful for other problems. First, we study the general
Gaussian observation model with parameterized mean or parameterized
covariance matrix. Indeed, one of the success of the Cram\'{e}r-Rao is that,
for this observation model, a closed-form expression of the Fisher
information matrix is available: this is the so-called Slepian-Bang formula 
\cite{Kay93}. Such kind of formulas have been less investigated in the
context of bounds tighter than the Cram\'{e}r-rao bound. Second, some
results are given in the multiple sources context without specifying the
structure of the steering matrix and of the noise covariance matrix.
Finally, these results are applied to the particular case of a single source
for two kinds of array geometries: the non-uniform linear array (elevation
only) and the planar (azimuth and elevation) array. Consequently, the aim of
this paper is also to provide a textbook of formulas which could be applied
in other fields. The Weiss-Weinstein bound is known to depend on parameters
called test points and other parameters generally denoted $s_{i}.$ One
particularity of this paper in comparison with the previous works on the
Weiss-Weinstein bound is that we do not use the assumption $s_{i}=1/2,$ $%
\forall i$.

This paper is organized as follows. Section \ref{Sec:Problem setup} is
devoted to the array processing observation model which will be used in the
paper. In Section \ref{Sec: WWB-generalities}, a short background on the
Weiss-Weinstein bound is presented and two general closed-form expressions
which will be the cornerstone for our array processing problems are derived.
In Section \ref{Sec: General application} we apply these general results to
the array processing problem without specifying the structure of the
steering matrix. In Section \ref{Sec: Specific application}, we study the
particular case of the non-uniform linear array and of the planar array for
which we provide both closed-form expressions of the bound. Some simulation
results are proposed in Section \ref{Sec: Simulation}. Finally, Section \ref%
{Sec: Conclusion} gives our conclusions.

\section{Problem setup\label{Sec:Problem setup}}

In this section, the general observation model generally used in array
signal processing is presented as well as the first different assumptions
used in the remain of the paper. Particularly, the so-called conditional and
unconditional source models are emphasized.

\subsection{Observations model}

We consider the classical scenario of an array with $M$ sensors which
receives $N$ complex bandpass signals $\mathbf{s}\left( t\right) =\left[
s_{1}\left( t\right) \ s_{2}\left( t\right) \cdots s_{N}\left( t\right) %
\right] ^{T}$. The output of the array is a $M\times 1$ complex vector $%
\mathbf{y}\left( t\right) $ which can be modelled as follows (see, \emph{e.g.%
}, \cite{VT02} or \cite{HLS93chap4})%
\begin{equation}
\mathbf{y}\left( t\right) =\mathbf{A}\left( \mathbf{\theta }\right) \mathbf{s%
}\left( t\right) +\mathbf{n}\left( t\right) ,\qquad t=1,\ldots ,T,
\label{eqn: array processing observation model}
\end{equation}%
where $T$ is the number of snapshots, where $\mathbf{\theta }=\left[ \theta
_{1}\ \theta _{2}\cdots \theta _{q}\right] ^{T}$ is an unknown parameter
vector of interest\footnote{%
Note that one source can be described by several parameters. Consequently, $%
q>N$ in general.}, where $\mathbf{A}\left( \mathbf{\theta }\right) $ is the
so-called $M\times N$ steering matrix of the array response to the sources,
and where the $M\times 1$ random vector $\mathbf{n}\left( t\right) $ is an
additive noise.

\subsection{Assumptions}

\begin{itemize}
\item The unknown parameters of interest are assumed to be random with an 
\emph{a priori} probability density function $p\left( \theta _{i}\right) ,$ $%
i=1,\ldots ,q$. These random parameters are assumed to be statistically
independent such that the \emph{a priori} joint probability density function
is $p\left( \mathbf{\theta }\right) =\overset{q}{\underset{i=1}{\dprod }}%
p\left( \theta _{i}\right) $. We also assume that the parameter space,
denoted $\Theta ,$ is a connected subset of $%
\mathbb{R}
^{q}$ (see \cite{BE08}).

\item The noise vector is assumed to be complex Gaussian, statistically
independent of the parameters, i.i.d., circular, with zero mean and known
covariance matrix $\mathbb{E}\left[ \mathbf{n}\left( t\right) \mathbf{n}%
^{H}\left( t\right) \right] =\mathbf{R}_{\mathbf{n}}$. This assumption will
be made more restrictive in Section \ref{Sec: Specific application} where it
will be assumed that $\mathbf{R}_{\mathbf{n}}=\sigma _{n}^{2}\mathbf{I.}$ In
any case, $\mathbf{R}_{\mathbf{n}}$ is assumed to be a full rank matrix.

\item The steering matrix $\mathbf{A}\left( \mathbf{\theta }\right) $ is
assumed such that the observation model is identifiable. From Section \ref%
{Sec: WWB-generalities} to Section \ref{Sec: General application}, the
structure of $\mathbf{A}\left( \mathbf{\theta }\right) $ is not specified in
order to obtain the more general results.

\item Concerning the source signals, two kinds of models have been
investigated in the literature (see, \emph{e.g.}, \cite{SN90a} or \cite%
{HLS93chap4}) and will be alternatively used in this paper.

\begin{itemize}
\item $\mathcal{M}_{1}$: \emph{Unconditional or stochastic model}: $\mathbf{s%
}(t)$ is assumed to be a complex circular random vector, i.i.d.,
statistically independent of the noise, Gaussian with zero-mean and known
covariance matrix $\mathbb{E}\left[ \mathbf{s}\left( t\right) \mathbf{s}%
^{H}\left( t\right) \right] =\mathbf{R}_{\mathbf{s}}$. Note that concerning
the previous results on the Cram\'{e}r-Rao bound available in the literature 
\cite{SN90a}, the covariance matrix $\mathbf{R}_{\mathbf{s}}$ is assumed to
be unknown. In this paper, we have made the simpler assumption that the
covariance matrix $\mathbf{R}_{\mathbf{s}}$ is known. These assumptions have
already been used for the calculation of bounds more complex than the Cram%
\'{e}r-Rao bound (see, \emph{e.g.}, \cite{XBR04}, \cite{BEV96b}, \cite{RM95}%
).

\item $\mathcal{M}_{2}$: \emph{Conditional or deterministic model}: $\forall
t$, $\mathbf{s}(t)$ is assumed to be deterministic known. Note that, under
the conditional model assumption, the signal waveforms can be assumed either
unknown or known. While the conditional observation model with unknown
waveforms seems more challenging, the conditional model with known waveforms
signals which will be used in this paper can be found in several
applications such as in mobile telecommunication and radar (see \textit{e.g.}
\cite{LC93},\cite{CM97},\cite{LHSV95},\cite{LV99}, and \cite{Cho04}).
\end{itemize}
\end{itemize}

\subsection{Likelihood of the observations}

Let $\mathbf{R}_{\mathbf{y}}=\mathbb{E}\left[ \mathbf{y}\left( t\right) 
\mathbf{y}^{H}\left( t\right) \right] $ be the covariance matrix of the
observation vector $\mathbf{y}\left( t\right) .$ According to the
aforementioned assumptions, it is easy to see that under $\mathcal{M}_{1}$,
the observations $\mathbf{y}\left( t\right) $ are distributed as a complex
circular Gaussian random vector with zero mean and covariance matrix $%
\mathbf{R}_{\mathbf{y}}(\mathbf{\theta })=\mathbf{A}(\mathbf{\theta })%
\mathbf{R}_{\mathbf{s}}\mathbf{A}^{H}(\mathbf{\theta })+\mathbf{R}_{\mathbf{n%
}}$ while under $\mathcal{M}_{2}$, the observations $\mathbf{y}\left(
t\right) $ are distributed as a complex circular Gaussian random vector with
mean $\mathbf{A}(\mathbf{\theta })\mathbf{s}\left( t\right) $ and covariance
matrix $\mathbf{R}_{\mathbf{y}}=\mathbf{R}_{\mathbf{n}}.$ Moreover, in both
case the observations are i.i.d..

Therefore, the likelihood, $p\left( \mathbf{Y};\mathbf{\theta }\right) ,$ of
the full observations matrix $\mathbf{Y}=\left[ \mathbf{y}\left( 1\right) \ 
\mathbf{y}\left( 2\right) \ \ldots \ \mathbf{y}\left( T\right) \right] $
under $\mathcal{M}_{1}$ is given by 
\begin{equation}
p\left( \mathbf{Y};\mathbf{\theta }\right) =\frac{1}{\pi ^{MT}\left\vert 
\mathbf{R}_{\mathbf{y}}(\mathbf{\theta })\right\vert ^{T}}\exp {\left( -%
\overset{T}{\underset{t=1}{\sum }}\mathbf{y}\left( t\right) ^{H}\mathbf{R}_{%
\mathbf{y}}^{-1}(\mathbf{\theta })\mathbf{y}\left( t\right) \right) },
\label{eqn: likelihood unconditional model}
\end{equation}%
where $\mathbf{R}_{\mathbf{y}}(\mathbf{\theta })=\mathbf{A}(\mathbf{\theta })%
\mathbf{R}_{\mathbf{s}}\mathbf{A}^{H}(\mathbf{\theta })+\mathbf{R}_{\mathbf{n%
}}$ and the likelihood under $\mathcal{M}_{2}$ is given by 
\begin{equation}
p\left( \mathbf{Y};\mathbf{\theta }\right) =\frac{1}{\pi ^{MT}\left\vert 
\mathbf{R}_{\mathbf{n}}\right\vert ^{T}}\exp {\left( -\overset{T}{\underset{%
t=1}{\sum }}\left( \mathbf{y}\left( t\right) -\mathbf{A}\left( \mathbf{%
\theta }\right) \mathbf{s}\left( t\right) \right) ^{H}\mathbf{R}_{\mathbf{n}%
}^{-1}\left( \mathbf{y}\left( t\right) -\mathbf{A}\left( \mathbf{\theta }%
\right) \mathbf{s}\left( t\right) \right) \right) }.
\end{equation}

\section{Weiss-Weinstein bound: Generalities\label{Sec: WWB-generalities}}

In this Section, we first remind to the reader the structure of the
Weiss-Weinstein bound on the mean square error and the assumptions used to
compute this bound. Second, a general result about the Gaussian observation
model with parameterized mean or parameterized covariance matrix, which, to
the best of our knowledge, does not appear in the literature is presented.
This result will be useful to study both the unconditional model $\mathcal{M}%
_{1}$ and the conditional model $\mathcal{M}_{2}$ in the next Section.

\subsection{Background}

The Weiss-Weinstein bound for a $q\times 1$ real parameter vector $\mathbf{%
\theta }$ is a $q\times q$ matrix denoted $\mathbf{WWB}$ and is given as
follows \cite{WW88} 
\begin{equation}
\mathbf{WWB}=\mathbf{HG}^{-1}\mathbf{H}^{T},
\end{equation}%
where the $q\times q$ matrix $\mathbf{H=}\left[ \mathbf{h}_{1}\text{ }%
\mathbf{h}_{2}\ldots \mathbf{h}_{q}\right] $ contains the so-called
test-points $\mathbf{h}_{i},$ $i=1,\ldots ,q$ such that $\mathbf{\theta +h}%
_{i}\in \Theta $ $\forall \mathbf{h}_{i}$. The $k,l-$element of the $q\times
q$ matrix $\mathbf{G}$ is given by%
\begin{equation}
\left\{ \mathbf{G}\right\} _{k,l}=\frac{\mathbb{E}\left[ \left(
L^{s_{k}}\left( \mathbf{Y};\mathbf{\theta +h}_{k},\mathbf{\theta }\right)
-L^{1-s_{k}}\left( \mathbf{Y};\mathbf{\theta -h}_{k},\mathbf{\theta }\right)
\right) \left( L^{s_{l}}\left( \mathbf{Y};\mathbf{\theta +h}_{l},\mathbf{%
\theta }\right) -L^{1-s_{l}}\left( \mathbf{Y};\mathbf{\theta -h}_{l},\mathbf{%
\theta }\right) \right) \right] }{\mathbb{E}\left[ L^{s_{k}}\left( \mathbf{Y}%
;\mathbf{\theta +h}_{k},\mathbf{\theta }\right) \right] \mathbb{E}\left[
L^{s_{l}}\left( \mathbf{Y};\mathbf{\theta +h}_{l},\mathbf{\theta }\right) %
\right] },  \label{eqn: element of G}
\end{equation}%
where the expectations are taken over the joint probability density function 
$p\left( \mathbf{Y,\theta }\right) $ and where the function $L\left( \mathbf{%
Y};\mathbf{\theta +h}_{i},\mathbf{\theta }\right) $ is defined by $L\left( 
\mathbf{Y};\mathbf{\theta +h}_{i},\mathbf{\theta }\right) =\frac{p\left( 
\mathbf{Y,\theta +h}_{i}\right) }{p\left( \mathbf{Y,\theta }\right) }$. The
elements $s_{i}$ are such that $s_{i}\in \left[ 0,1\right] $, $i=1,\ldots ,q$%
.

Note that we have the following order relation \cite{WW88}%
\begin{equation}
\mathbf{Cov}\left( \mathbf{\hat{\theta}}\right) =\mathbb{E}\left[ \left( 
\mathbf{\hat{\theta}-\theta }\right) \left( \mathbf{\hat{\theta}-\theta }%
\right) ^{T}\right] \succeq \mathbf{WWB,}
\end{equation}%
where $\mathbf{A\succeq B}$ means that the matrix $\mathbf{A}-\mathbf{B}$ is
a semi-positive definite matrix and where $\mathbf{Cov}\left( \mathbf{\hat{%
\theta}}\right) $ is the global (the expectation is taken over the joint pdf 
$p\left( \mathbf{Y,\theta }\right) $) mean square error of any estimator $%
\mathbf{\hat{\theta}}$ of the parameter vector $\mathbf{\theta }$. Finally,
in order to obtain a tight bound, one has to maximize $\mathbf{WWB}$ over
the test-points $\mathbf{h}_{i}$ and $s_{i}$ $i=1,\ldots ,q.$ Note that this
maximization can be done by using the trace of $\mathbf{HG}^{-1}\mathbf{H}%
^{T}$ or with respect to the Loewner partial ordering \cite{LRNM10}. In this
paper we will use the trace of $\mathbf{HG}^{-1}\mathbf{H}^{T}$ which is
enough to obtain tight results.

\subsection{A general result on the Weiss-Weinstein bound and its
application to the Gaussian observation models\label{Sec: General result}}

An analytical result on the Weiss-Weinstein bound which will be useful in
the following derivations and which could be useful for other problems is
derived in this part. Note that this result is independent of the parameter
vector size $q$ and of the considered observation model.

Let us denote $\Omega $ the observation space. By rewriting the elements of
matrix $\mathbf{G}$ (see Eqn. (\ref{eqn: element of G})) involved in the
Weiss-Weinstein bound, one obtains for the numerator denoted $N_{\left\{ 
\mathbf{G}\right\} _{k,l}},$%
\begin{eqnarray}
N_{\left\{ \mathbf{G}\right\} _{k,l}} &=&\mathbb{E}\left[ \left(
L^{s_{k}}\left( \mathbf{Y};\mathbf{\theta +h}_{k},\mathbf{\theta }\right)
-L^{1-s_{k}}\left( \mathbf{Y};\mathbf{\theta -h}_{k},\mathbf{\theta }\right)
\right) \left( L^{s_{l}}\left( \mathbf{Y};\mathbf{\theta +h}_{l},\mathbf{%
\theta }\right) -L^{1-s_{l}}\left( \mathbf{Y};\mathbf{\theta -h}_{l},\mathbf{%
\theta }\right) \right) \right]  \notag \\
&=&\int_{\Theta }\int_{\Omega }\frac{p^{s_{k}}\left( \mathbf{Y,\theta +h}%
_{k}\right) p^{s_{l}}\left( \mathbf{Y,\theta +h}_{l}\right) }{%
p^{s_{k}+s_{l}-1}\left( \mathbf{Y,\theta }\right) }d\mathbf{Y}d\mathbf{%
\theta +}\int_{\Theta }\int_{\Omega }\frac{p^{1-s_{k}}\left( \mathbf{%
Y,\theta -h}_{k}\right) p^{1-s_{l}}\left( \mathbf{Y,\theta -h}_{l}\right) }{%
p^{1-s_{k}-s_{l}}\left( \mathbf{Y,\theta }\right) }d\mathbf{Y}d\mathbf{%
\theta }  \notag \\
&&\hspace{-1cm}\mathbf{-}\int_{\Theta }\int_{\Omega }\frac{p^{s_{k}}\left( 
\mathbf{Y,\theta +h}_{k}\right) p^{1-s_{l}}\left( \mathbf{Y,\theta -h}%
_{l}\right) }{p^{s_{k}-s_{l}}\left( \mathbf{Y,\theta }\right) }d\mathbf{Y}d%
\mathbf{\theta -}\int_{\Theta }\int_{\Omega }\frac{p^{1-s_{k}}\left( \mathbf{%
Y,\theta -h}_{k}\right) p^{s_{l}}\left( \mathbf{Y,\theta +h}_{l}\right) }{%
p^{s_{l}-s_{k}}\left( \mathbf{Y,\theta }\right) }d\mathbf{Y}d\mathbf{\theta ,%
}
\end{eqnarray}%
and for the denominator denoted $D_{\left\{ \mathbf{G}\right\} _{k,l}},$%
\begin{eqnarray}
D_{\left\{ \mathbf{G}\right\} _{k,l}} &=&\mathbb{E}\left[ L^{s_{k}}\left( 
\mathbf{Y};\mathbf{\theta +h}_{k},\mathbf{\theta }\right) \right] \mathbb{E}%
\left[ L^{s_{l}}\left( \mathbf{Y};\mathbf{\theta +h}_{l},\mathbf{\theta }%
\right) \right]  \notag \\
&=&\int_{\Theta }\int_{\Omega }\frac{p^{s_{k}}\left( \mathbf{Y,\theta +h}%
_{k}\right) }{p^{s_{k}-1}\left( \mathbf{Y,\theta }\right) }d\mathbf{Y}d%
\mathbf{\theta }\int_{\Theta }\int_{\Omega }\frac{p^{s_{l}}\left( \mathbf{%
Y,\theta +h}_{l}\right) }{p^{s_{l}-1}\left( \mathbf{Y,\theta }\right) }d%
\mathbf{Y}d\mathbf{\theta .}
\end{eqnarray}

Let us now define a function $\eta \left( \alpha ,\beta ,\mathbf{u},\mathbf{v%
}\right) $ as%
\begin{equation}
\eta \left( \alpha ,\beta ,\mathbf{u},\mathbf{v}\right) =\int_{\Theta
}\int_{\Omega }\frac{p^{\alpha }\left( \mathbf{Y,\theta +u}\right) p^{\beta
}\left( \mathbf{Y,\theta +v}\right) }{p^{\alpha +\beta -1}\left( \mathbf{%
Y,\theta }\right) }d\mathbf{Y}d\mathbf{\theta ,}
\label{eqn: general function eta}
\end{equation}%
where $\left( \alpha ,\beta \right) \in \left[ 0,1\right] ^{2}$ and where $%
\left( \mathbf{u},\mathbf{v}\right) $ are two $q\times 1$ vectors such that $%
\mathbf{\theta +u}\in \Theta $ and $\mathbf{\theta +v}\in \Theta $. By
identification, it is easy to see that%
\begin{eqnarray}
\left\{ \mathbf{G}\right\} _{k,l} &=&  \notag \\
&&\hspace{-1cm}\frac{\eta \left( s_{k},s_{l},\mathbf{h}_{k},\mathbf{h}%
_{l}\right) +\eta \left( 1-s_{k},1-s_{l},\mathbf{-h}_{k},\mathbf{-h}%
_{l}\right) -\eta \left( s_{k},1-s_{l},\mathbf{h}_{k},\mathbf{-h}_{l}\right)
-\eta \left( 1-s_{k},s_{l},\mathbf{-h}_{k},\mathbf{h}_{l}\right) }{\eta
\left( s_{k},0,\mathbf{h}_{k},\mathbf{0}\right) \eta \left( 0,s_{l},\mathbf{0%
},\mathbf{h}_{l}\right) }.  \label{eqn: element of G rewritten}
\end{eqnarray}

Note that we choose the arbitrary notation $D_{\left\{ \mathbf{G}\right\}
_{k,l}}=\eta \left( s_{k},0,\mathbf{h}_{k},\mathbf{0}\right) \eta \left(
0,s_{l},\mathbf{0},\mathbf{h}_{l}\right) $ for the denominator. The notation 
$D_{\left\{ \mathbf{G}\right\} _{k,l}}=\eta \left( s_{k},1,\mathbf{h}_{k},%
\mathbf{0}\right) \eta \left( 1,s_{l},\mathbf{0},\mathbf{h}_{l}\right) $ or,
even, $D_{\left\{ \mathbf{G}\right\} _{k,l}}=\eta \left( s_{k},0,\mathbf{h}%
_{k},\mathbf{v}\right) \eta \left( 0,s_{l},\mathbf{u},\mathbf{h}_{l}\right) $
will lead to the same result.

With Eqn. (\ref{eqn: element of G rewritten}), it is clear that the
knowledge of $\eta \left( \alpha ,\beta ,\mathbf{u},\mathbf{v}\right) $ for
a particular problem leads to the Weiss-Weinstein bound (without the
maximization procedure over the test-points and over the parameters $s_{i}$%
). Surprisingly, this simple expression is given in \cite{WW88} only for $%
s_{i}=\frac{1}{2},$ $\forall i$ and not for the general case.

Let us now detail this function $\eta \left( \alpha ,\beta ,\mathbf{u},%
\mathbf{v}\right) $. The function $\eta \left( \alpha ,\beta ,\mathbf{u},%
\mathbf{v}\right) $ can be rewritten as%
\begin{eqnarray}
\eta \left( \alpha ,\beta ,\mathbf{u},\mathbf{v}\right) &=&\int_{\Theta }%
\frac{p^{\alpha }\left( \mathbf{\theta +u}\right) p^{\beta }\left( \mathbf{%
\theta +v}\right) }{p^{\alpha +\beta -1}\left( \mathbf{\theta }\right) }%
\int_{\Omega }\frac{p^{\alpha }\left( \mathbf{Y;\theta +u}\right) p^{\beta
}\left( \mathbf{Y;\theta +v}\right) }{p^{\alpha +\beta -1}\left( \mathbf{%
Y;\theta }\right) }d\mathbf{Y}d\mathbf{\theta }  \notag \\
&=&\int_{\Theta }\acute{\eta}_{\mathbf{\theta }}\left( \alpha ,\beta ,%
\mathbf{u},\mathbf{v}\right) \frac{p^{\alpha }\left( \mathbf{\theta +u}%
\right) p^{\beta }\left( \mathbf{\theta +v}\right) }{p^{\alpha +\beta
-1}\left( \mathbf{\theta }\right) }d\mathbf{\theta ,}
\label{eqn: general function eta rewritten}
\end{eqnarray}%
where we define%
\begin{equation}
\acute{\eta}_{\mathbf{\theta }}\left( \alpha ,\beta ,\mathbf{u},\mathbf{%
v,\theta }\right) =\int_{\Omega }\frac{p^{\alpha }\left( \mathbf{Y;\theta +u}%
\right) p^{\beta }\left( \mathbf{Y;\theta +v}\right) }{p^{\alpha +\beta
-1}\left( \mathbf{Y;\theta }\right) }d\mathbf{Y.}
\label{eqn: general function eta prime}
\end{equation}

Our aim is to give the most general result. Consequently, we will focus only
on $\acute{\eta}_{\mathbf{\theta }}\left( \alpha ,\beta ,\mathbf{u},\mathbf{v%
}\right) $ since the \emph{a priori} probability density function depends on
the considered problem.

An important remark pointed out in \cite{BE08} is that the integration for
the parameter space is with respect to the region $\left\{ \mathbf{\theta }%
:p\left( \mathbf{\theta }\right) >0\right\} .$ However, since the functions
being integrated are $p\left( \mathbf{\theta }\right) ,$ $p\left( \mathbf{%
\theta +u}\right) ,$ and $p\left( \mathbf{\theta +v}\right) ,$ then the
actual region of integration (where all the functions are positive) is the
intersection of three regions, $\left\{ \mathbf{\theta }:p\left( \mathbf{%
\theta }\right) >0\right\} \cap \left\{ \mathbf{\theta }:p\left( \mathbf{%
\theta +u}\right) >0\right\} \cap \left\{ \mathbf{\theta }:p\left( \mathbf{%
\theta +v}\right) >0\right\} .$ Note that, in order to simplify the notation
we only use $\Theta $ throughout this paper but this remark will be useful
and explictely specified in Section \ref{Sec: Analysis of eta with uniform
prior}.

\subsubsection{Gaussian observation model with parameterized covariance
matrix}

One calls (circular, i.i.d.) Gaussian observation model with parameterized
covariance matrix, a model such that the observations $\mathbf{y}\left(
t\right) \sim \mathcal{CN}\left( \mathbf{0},\mathbf{R}_{\mathbf{y}}\left( 
\mathbf{\theta }\right) \right) $ where $\mathbf{\theta }$ are the
parameters of interest. Note that $\mathcal{M}_{1}$ is a special case of
this model since the parameters of interest appear only in the covariance
matrix of the observations which has the following particular structure $%
\mathbf{R}_{\mathbf{y}}(\mathbf{\theta })=\mathbf{A}(\mathbf{\theta })%
\mathbf{R}_{\mathbf{s}}\mathbf{A}^{H}(\mathbf{\theta })+\mathbf{R}_{\mathbf{n%
}}$. The closed-form expression of $\acute{\eta}_{\mathbf{\theta }}\left(
\alpha ,\beta ,\mathbf{u},\mathbf{v}\right) $ is given by: 
\begin{equation}
\acute{\eta}_{\mathbf{\theta }}\left( \alpha ,\beta ,\mathbf{u},\mathbf{v}%
\right) =\frac{\left\vert \mathbf{R}_{\mathbf{y}}(\mathbf{\theta }%
)\right\vert ^{T(\alpha +\beta -1)}}{\left\vert \mathbf{R}_{\mathbf{y}}(%
\mathbf{\theta +u})\right\vert ^{T\alpha }\left\vert \mathbf{R}_{\mathbf{y}}(%
\mathbf{\theta +v})\right\vert ^{T\beta }\left\vert \alpha \mathbf{R}_{%
\mathbf{y}}^{-1}(\mathbf{\theta +u})+\beta \mathbf{R}_{\mathbf{y}}^{-1}(%
\mathbf{\theta +v})-(\alpha +\beta -1)\mathbf{R}_{\mathbf{y}}^{-1}(\mathbf{%
\theta })\right\vert ^{T}}.
\label{eqn: eta prime gaussian model with parameterized covariance}
\end{equation}%
The proof is given in Appendix \ref{Sec: Appendix A}. Note that, similar
expressions are given in \cite{BEV96a} (Eqn. (B.15)) and \cite{VT01} (p. 67,
Eqn. (52)) for the particular case where $\alpha =s$ and $\beta =1-s.$

\subsubsection{Gaussian observation model with parameterized mean}

One calls (circular, i.i.d.) Gaussian observation model with parameterized
mean, a model such that the observations $\mathbf{y}\left( t\right) \sim 
\mathcal{CN}\left( \mathbf{f}\left( \mathbf{\theta }\right) ,\mathbf{R}_{%
\mathbf{y}}\right) $ where $\mathbf{\theta }$ are the parameters of
interest. Note that $\mathcal{M}_{2}$ is a special case of this model since
the parameters of interest appear only in the mean of the observations which
has the following particular structure $\mathbf{f}_{t}\left( \mathbf{\theta }%
\right) =\mathbf{A}(\mathbf{\theta })\mathbf{s}\left( t\right) $ (and $%
\mathbf{R}_{\mathbf{y}}=\mathbf{R}_{\mathbf{n}}$). The closed-form
expression of $\acute{\eta}_{\mathbf{\theta }}\left( \alpha ,\beta ,\mathbf{u%
},\mathbf{v}\right) $ is given in this case by

\begin{eqnarray}
\ln \acute{\eta}_{\mathbf{\theta }}\left( \alpha ,\beta ,\mathbf{u},\mathbf{v%
}\right)  &=&-\overset{T}{\underset{t=1}{\sum }}\alpha \left( 1-\alpha
\right) \mathbf{f}_{t}^{H}\left( \mathbf{\theta +u}\right) \mathbf{R}_{%
\mathbf{y}}^{-1}\mathbf{f}_{t}\left( \mathbf{\theta +u}\right) \mathbf{+}%
\beta \left( 1-\beta \right) \mathbf{f}_{t}^{H}\left( \mathbf{\theta +v}%
\right) \mathbf{R}_{\mathbf{y}}^{-1}\mathbf{f}_{t}\left( \mathbf{\theta +v}%
\right)   \notag \\
&&\mathbf{+}\left( 1-\alpha -\beta \right) \left( \alpha +\beta \right) 
\mathbf{f}_{t}^{H}\left( \mathbf{\theta }\right) \mathbf{R}_{\mathbf{y}}^{-1}%
\mathbf{f}_{t}\left( \mathbf{\theta }\right) -2\func{Re}\left\{ \alpha \beta 
\mathbf{f}_{t}^{H}\left( \mathbf{\theta +u}\right) \mathbf{R}_{\mathbf{y}%
}^{-1}\mathbf{f}_{t}\left( \mathbf{\theta +v}\right) \right.   \notag \\
&&\left. +\alpha \left( 1-\alpha -\beta \right) \mathbf{f}_{t}^{H}\left( 
\mathbf{\theta +u}\right) \mathbf{R}_{\mathbf{y}}^{-1}\mathbf{f}_{t}\left( 
\mathbf{\theta }\right) +\beta \left( 1-\alpha -\beta \right) \mathbf{f}%
_{t}^{H}\left( \mathbf{\theta +v}\right) \mathbf{R}_{\mathbf{y}}^{-1}\mathbf{%
f}_{t}\left( \mathbf{\theta }\right) \right\} ,
\end{eqnarray}%
or equivalently by%
\begin{eqnarray}
\ln \acute{\eta}_{\mathbf{\theta }}\left( \alpha ,\beta ,\mathbf{u},\mathbf{v%
}\right)  &=&-\overset{T}{\underset{t=1}{\sum }}\alpha \left( 1-\alpha
-\beta \right) \left\Vert \mathbf{R}_{\mathbf{y}}^{-1/2}\left( \mathbf{f}%
_{t}\left( \mathbf{\theta +u}\right) -\mathbf{f}_{t}\left( \mathbf{\theta }%
\right) \right) \right\Vert ^{2}+\alpha \beta \left\Vert \mathbf{R}_{\mathbf{%
y}}^{-1/2}\left( \mathbf{f}_{t}\left( \mathbf{\theta +u}\right) -\mathbf{f}%
_{t}\left( \mathbf{\theta +v}\right) \right) \right\Vert ^{2}  \notag \\
&&+\beta \left( 1-\alpha -\beta \right) \left\Vert \mathbf{R}_{\mathbf{y}%
}^{-1/2}\left( \mathbf{f}_{t}\left( \mathbf{\theta +v}\right) -\mathbf{f}%
_{t}\left( \mathbf{\theta }\right) \right) \right\Vert ^{2}.
\label{eqn: eta prime gaussian model with parameterized mean}
\end{eqnarray}

The details are given in Appendix \ref{Sec: Appendix B}.

\section{General application to array processing\label{Sec: General
application}}

In the previous Section, it has been shown that the Weiss-Weinstein bound
computation (or, at least, the matrix $\mathbf{G}$ computation) is reduced
to the knowledge of the function $\eta \left( \alpha ,\beta ,\mathbf{u},%
\mathbf{v}\right) $ given by Eqn. (\ref{eqn: general function eta}). As one
can see in Eqn. (\ref{eqn: element of G rewritten}), the elements of the
matrix $\mathbf{G}$ depend on $\eta \left( \alpha ,\beta ,\mathbf{u},\mathbf{%
v}\right) $ for particular values of $\alpha ,$ $\beta ,$ $\mathbf{u},$ and $%
\mathbf{v.}$ Consequently, the goal of this Section is to detail these
particular functions for our model given by Eqn. (\ref{eqn: array processing
observation model}). Since Eqn. (\ref{eqn: general function eta}) can be
decomposed into a \emph{deterministic part} (in the sense where $\acute{\eta}%
_{\mathbf{\theta }}\left( \alpha ,\beta ,\mathbf{u},\mathbf{v}\right) $ (see
Eqn. (\ref{eqn: general function eta prime})) only depends on the likelihood
function) and a \emph{Bayesian part} (when we have to integrate $\acute{\eta}%
_{\mathbf{\theta }}\left( \alpha ,\beta ,\mathbf{u},\mathbf{v}\right) $ over
the \emph{a priori} probability density function of the parameters), we will
first focus on the particular functions $\acute{\eta}_{\mathbf{\theta }%
}\left( \alpha ,\beta ,\mathbf{u},\mathbf{v}\right) $ by using the results
of the previous Section on the Gaussian observation model with parameterized
mean or covariance matrix. Second, we will detail the passage from $\acute{%
\eta}_{\mathbf{\theta }}\left( \alpha ,\beta ,\mathbf{u},\mathbf{v}\right) $
to $\eta \left( \alpha ,\beta ,\mathbf{u},\mathbf{v}\right) $ in the
particular case where $p\left( \theta _{i}\right) $ is a uniform probability
density function $\forall i$. Another result will also be given in the case
of a Gaussian prior.

\subsection{Analysis of $\acute{\protect\eta}_{\mathbf{\protect\theta }%
}\left( \protect\alpha ,\protect\beta ,\mathbf{u},\mathbf{v}\right) $}

We will now detail the particular functions $\acute{\eta}_{\mathbf{\theta }%
}\left( \alpha ,\beta ,\mathbf{u},\mathbf{v}\right) $ involved in the
different elements of $\left\{ \mathbf{G}\right\} _{k,l},$ $k,l\in \left\{
1,q\right\} ^{2}$ for both models $\mathcal{M}_{1}$ and $\mathcal{M}_{2}.$

\subsubsection{Unconditional observation model $\mathcal{M}_{1}$}

Under the unconditional model $\mathcal{M}_{1}$, by using Eqn. (\ref{eqn:
eta prime gaussian model with parameterized covariance}), one obtains
straightforwardly the functions $\acute{\eta}_{\mathbf{\theta }}\left(
\alpha ,\beta ,\mathbf{u},\mathbf{v}\right) $ involved in the elements $%
\left\{ \mathbf{G}\right\} _{k,l}=\left\{ \mathbf{G}\right\} _{l,k}$

\begin{equation}
\left\{ 
\begin{array}{l}
\acute{\eta}_{\mathbf{\theta }}(s_{k},s_{l},\mathbf{h}_{k},\mathbf{h}_{l})=%
\frac{\left\vert \mathbf{R}_{\mathbf{y}}(\mathbf{\theta })\right\vert
^{T\left( s_{k}+s_{l}-1\right) }}{\left\vert \mathbf{R}_{\mathbf{y}}(\mathbf{%
\theta }+\mathbf{h}_{k})\right\vert ^{Ts_{k}}\left\vert \mathbf{R}_{\mathbf{y%
}}(\mathbf{\theta }+\mathbf{h}_{l})\right\vert ^{Ts_{l}}\left\vert s_{k}%
\mathbf{R}_{\mathbf{y}}^{-1}(\mathbf{\theta }+\mathbf{h}_{k})+s_{l}\mathbf{R}%
_{\mathbf{y}}^{-1}(\mathbf{\theta }+\mathbf{h}_{l})-(s_{k}+s_{l}-1)\mathbf{R}%
_{\mathbf{y}}^{-1}(\mathbf{\theta })\right\vert ^{T}}, \\ 
\acute{\eta}_{\mathbf{\theta }}(1-s_{k},1-s_{l},-\mathbf{h}_{k},-\mathbf{h}%
_{l})=\frac{\left\vert \mathbf{R}_{\mathbf{y}}(\mathbf{\theta })\right\vert
^{T\left( 1-s_{k}-s_{l}\right) }\left\vert \mathbf{R}_{\mathbf{y}}(\mathbf{%
\theta }-\mathbf{h}_{k})\right\vert ^{T\left( s_{k}-1\right) }\left\vert 
\mathbf{R}_{\mathbf{y}}(\mathbf{\theta }-\mathbf{h}_{l})\right\vert
^{T\left( s_{l}-1\right) }}{\left\vert (1-s_{k})\mathbf{R}_{\mathbf{y}}^{-1}(%
\mathbf{\theta }-\mathbf{h}_{k})+(1-s_{l})\mathbf{R}_{\mathbf{y}}^{-1}(%
\mathbf{\theta }-\mathbf{h}_{l})-(1-s_{k}-s_{l})\mathbf{R}_{\mathbf{y}}^{-1}(%
\mathbf{\theta })\right\vert ^{T}}, \\ 
\acute{\eta}_{\mathbf{\theta }}(s_{k},1-s_{l},\mathbf{h}_{k},-\mathbf{h}%
_{l})=\frac{\left\vert \mathbf{R}_{\mathbf{y}}(\mathbf{\theta })\right\vert
^{T\left( s_{k}-s_{l}\right) }\left\vert \mathbf{R}_{\mathbf{y}}(\mathbf{%
\theta }-\mathbf{h}_{l})\right\vert ^{T\left( s_{l}-1\right) }}{\left\vert 
\mathbf{R}_{\mathbf{y}}(\mathbf{\theta }+\mathbf{h}_{k})\right\vert
^{Ts_{k}}\left\vert s_{k}\mathbf{R}_{\mathbf{y}}^{-1}(\mathbf{\theta }+%
\mathbf{h}_{k})+(1-s_{l})\mathbf{R}_{\mathbf{y}}^{-1}(\mathbf{\theta }-%
\mathbf{h}_{l})-(s_{k}-s_{l})\mathbf{R}_{\mathbf{y}}^{-1}(\mathbf{\theta }%
)\right\vert ^{T}}, \\ 
\acute{\eta}_{\mathbf{\theta }}(1-s_{k},s_{l},-\mathbf{h}_{k},\mathbf{h}%
_{l})=\frac{\left\vert \mathbf{R}_{\mathbf{y}}(\mathbf{\theta })\right\vert
^{T\left( s_{l}-s_{k}\right) }\left\vert \mathbf{R}_{\mathbf{y}}(\mathbf{%
\theta }-\mathbf{h}_{k})\right\vert ^{T\left( s_{k}-1\right) }}{\left\vert 
\mathbf{R}_{\mathbf{y}}(\mathbf{\theta }+\mathbf{h}_{l})\right\vert
^{Ts_{l}}\left\vert (1-s_{k})\mathbf{R}_{\mathbf{y}}^{-1}(\mathbf{\theta }-%
\mathbf{h}_{k})+s_{l}\mathbf{R}_{\mathbf{y}}^{-1}(\mathbf{\theta }+\mathbf{h}%
_{l})-(s_{l}-s_{k})\mathbf{R}_{\mathbf{y}}^{-1}(\mathbf{\theta })\right\vert
^{T}}, \\ 
\acute{\eta}_{\mathbf{\theta }}(s_{k},0,\mathbf{h}_{k},\mathbf{0})=\frac{%
\left\vert \mathbf{R}_{\mathbf{y}}(\mathbf{\theta })\right\vert ^{T\left(
s_{k}-1\right) }}{\left\vert \mathbf{R}_{\mathbf{y}}(\mathbf{\theta }+%
\mathbf{h}_{k})\right\vert ^{Ts_{k}}\left\vert s_{k}\mathbf{R}_{\mathbf{y}%
}^{-1}(\mathbf{\theta }+\mathbf{h}_{k})-(s_{k}-1)\mathbf{R}_{\mathbf{y}%
}^{-1}(\mathbf{\theta })\right\vert ^{T}}, \\ 
\acute{\eta}_{\mathbf{\theta }}(0,s_{l},\mathbf{0},\mathbf{h}_{l})=\frac{%
\left\vert \mathbf{R}_{\mathbf{y}}(\mathbf{\theta })\right\vert ^{T\left(
s_{l}-1\right) }}{\left\vert \mathbf{R}_{\mathbf{y}}(\mathbf{\theta }+%
\mathbf{h}_{l})\right\vert ^{Ts_{l}}\left\vert s_{l}\mathbf{R}_{\mathbf{y}%
}^{-1}(\mathbf{\theta }+\mathbf{h}_{l})-(s_{l}-1)\mathbf{R}_{\mathbf{y}%
}^{-1}(\mathbf{\theta })\right\vert ^{T}}.%
\end{array}%
\right.  \label{eqn: set of eta prime function unconditional model}
\end{equation}

The diagonal elements of $\mathbf{G}$ are obtained by letting $k=l$ in the
above equations.

\subsubsection{Conditional observation model $\mathcal{M}_{2}$}

Under the conditional model $\mathcal{M}_{2}$, by using Eqn. (\ref{eqn: eta
prime gaussian model with parameterized mean}) with $\mathbf{f}_{t}\left( 
\mathbf{\theta }\right) =\mathbf{A}\left( \mathbf{\theta }\right) \mathbf{s}%
\left( t\right) $ and $\mathbf{R}_{\mathbf{y}}=\mathbf{R}_{\mathbf{n}}$ one
obtains straightforwardly the functions $\acute{\eta}_{\mathbf{\theta }%
}\left( \alpha ,\beta ,\mathbf{u},\mathbf{v}\right) $ involved in the
elements $\left\{ \mathbf{G}\right\} _{k,l}=\left\{ \mathbf{G}\right\} _{l,k}
$

\begin{equation}
\left\{ 
\begin{array}{l}
\ln \acute{\eta}_{\mathbf{\theta }}\left( s_{k},s_{l},\mathbf{h}_{k},\mathbf{%
h}_{l}\right) =s_{k}\left( s_{k}+s_{l}-1\right) \zeta _{\mathbf{\theta }%
}\left( \mathbf{h}_{k}\mathbf{,0}\right) +s_{l}\left( s_{k}+s_{l}-1\right)
\zeta _{\mathbf{\theta }}\left( \mathbf{h}_{l}\mathbf{,0}\right)
-s_{k}s_{l}\zeta _{\mathbf{\theta }}\left( \mathbf{h}_{k}\mathbf{,h}%
_{l}\right) , \\ 
\ln \acute{\eta}_{\mathbf{\theta }}\left( 1-s_{k},1-s_{l},\mathbf{-h}_{k},%
\mathbf{-h}_{l}\right) =\left( s_{k}-1\right) \left( s_{k}+s_{l}-1\right)
\zeta _{\mathbf{\theta }}\left( -\mathbf{h}_{k}\mathbf{,0}\right) +\left(
s_{l}-1\right) \left( s_{k}+s_{l}-1\right) \zeta _{\mathbf{\theta }}\left( -%
\mathbf{h}_{l}\mathbf{,0}\right) \\ 
\hspace{5cm}-\left( 1-s_{k}\right) \left( 1-s_{l}\right) \zeta _{\mathbf{%
\theta }}\left( -\mathbf{h}_{k}\mathbf{,-h}_{l}\right) , \\ 
\ln \acute{\eta}_{\mathbf{\theta }}\left( s_{k},1-s_{l},\mathbf{h}_{k},%
\mathbf{-h}_{l}\right) =s_{k}\left( s_{k}-s_{l}\right) \zeta _{\mathbf{%
\theta }}\left( \mathbf{h}_{k}\mathbf{,0}\right) +\left( 1-s_{l}\right)
\left( s_{k}-s_{l}\right) \zeta _{\mathbf{\theta }}\left( -\mathbf{h}_{l}%
\mathbf{,0}\right) +s_{k}\left( s_{l}-1\right) \zeta _{\mathbf{\theta }%
}\left( \mathbf{h}_{k}\mathbf{,-h}_{l}\right) , \\ 
\ln \acute{\eta}_{\mathbf{\theta }}\left( 1-s_{k},s_{l},\mathbf{-h}_{k},%
\mathbf{h}_{l}\right) =\left( s_{k}-1\right) \left( s_{k}-s_{l}\right) \zeta
_{\mathbf{\theta }}\left( -\mathbf{h}_{k}\mathbf{,0}\right) +s_{l}\left(
s_{l}-s_{k}\right) \zeta _{\mathbf{\theta }}\left( \mathbf{h}_{l}\mathbf{,0}%
\right) +\left( s_{k}-1\right) s_{l}\zeta _{\mathbf{\theta }}\left( -\mathbf{%
h}_{k}\mathbf{,h}_{l}\right) , \\ 
\ln \acute{\eta}_{\mathbf{\theta }}\left( s_{k},0,\mathbf{h}_{k},\mathbf{0}%
\right) =s_{k}\left( s_{k}-1\right) \zeta _{\mathbf{\theta }}\left( \mathbf{h%
}_{k}\mathbf{,0}\right) , \\ 
\ln \acute{\eta}_{\mathbf{\theta }}\left( 0,s_{l},\mathbf{0},\mathbf{h}%
_{l}\right) =s_{l}\left( s_{l}-1\right) \zeta _{\mathbf{\theta }}\left( 
\mathbf{h}_{l}\mathbf{,0}\right) ,%
\end{array}%
\right.  \label{eqn: set of eta prime function conditional model}
\end{equation}%
where we define%
\begin{equation}
\zeta _{\mathbf{\theta }}\left( \mathbf{\mu ,\rho }\right) =\overset{T}{%
\underset{t=1}{\sum }}\left\Vert \mathbf{R}_{\mathbf{n}}^{-1/2}\left( 
\mathbf{A}\left( \mathbf{\theta +\mu }\right) -\mathbf{A}\left( \mathbf{%
\theta +\rho }\right) \right) \mathbf{s}\left( t\right) \right\Vert ^{2}.
\label{eqn: general function zeta}
\end{equation}

The diagonal elements of $\mathbf{G}$ are obtained by letting $k=l$ in the
above equations. Note that, since we are working on matrix $\mathbf{G}$, all
the previously proposed results are made whatever the number of test-points.

\subsection{Analysis of $\protect\eta \left( \protect\alpha ,\protect\beta ,%
\mathbf{u},\mathbf{v}\right) $ with a uniform prior\label{Sec: Analysis of
eta with uniform prior}}

Of course, the analysis of $\eta \left( \alpha ,\beta ,\mathbf{u},\mathbf{v}%
\right) $ given by Eqn. (\ref{eqn: general function eta rewritten}) can only
be conducted by specifying the \emph{a priori} probability density functions
of the parameters. Consequently, the results provided here are very
specific. However, note that, in general, this aspect is less emphasized in
the literature where most of the authors give results without specifying the
prior probability density functions and compute the rest of the bound
numerically (see e.g., \cite{XBR04}\cite{NVT94}\cite{Bel95}).

We assume that all the parameters $\theta _{i}$ have a uniform prior
distribution over the interval $[a_{i},b_{i}]$ and are statistically
independent. We will also assume one test-point per parameter, otherwise
there is no possibility to obtain (pseudo) closed-form expressions.
Consequently, the matrix $\mathbf{H}$ is such that%
\begin{equation}
\mathbf{H}=Diag\left( \left[ h_{1}\ h_{2}\cdots h_{q}\right] \right) ,
\end{equation}%
and the vector $\mathbf{h}_{i},$ $i=1,\ldots ,q$, takes the value $h_{i}$ at
the $i^{th}$ row and zero elsewhere. So, in this analysis, the vector $%
\mathbf{u}$ takes the value $u_{i}$ at the $i^{th}$ row and zero elsewhere
and the vector $\mathbf{v}$ takes the value $v_{j}$ at the $j^{th}$ row and
zero elsewhere (of course, we can have $i=j$). Under these assumptions, $%
\eta \left( \alpha ,\beta ,\mathbf{u},\mathbf{v}\right) $ can be rewritten%
\footnote{%
In this case, one has to have a particular attention to the integration
domain as mentionned in Section \ref{Sec: General result}. It will not be
the case for the Gaussian prior since the support is $%
\mathbb{R}
.$} for $i\neq j$ 
\begin{eqnarray}
\eta \left( \alpha ,\beta ,\mathbf{u},\mathbf{v}\right)  &=&\int_{\Theta }%
\acute{\eta}_{\mathbf{\theta }}\left( \alpha ,\beta ,\mathbf{u},\mathbf{v}%
\right) \frac{p^{\alpha }\left( \theta _{i}+u_{i}\right) p^{\beta }\left(
\theta _{j}+v_{j}\right) p^{\beta }\left( \theta _{i}\right) p^{\alpha
}\left( \theta _{j}\right) }{p^{\alpha +\beta -1}\left( \theta _{i}\right)
p^{\alpha +\beta -1}\left( \theta _{j}\right) }\overset{q}{\underset{k\neq
i,k\neq j}{\underset{k=1}{\dprod }}}p\left( \theta _{k}\right) d\mathbf{%
\theta }  \notag \\
&=&\frac{1}{\overset{q}{\underset{k=1}{\dprod }}\left( b_{k}-a_{k}\right) }%
\int_{\Theta ^{q-2}}\int_{\Theta _{j}}\int_{\Theta _{i}}\acute{\eta}_{%
\mathbf{\theta }}\left( \alpha ,\beta ,\mathbf{u},\mathbf{v}\right) d\theta
_{i}d\theta _{j}d\left( \mathbf{\theta /}\left\{ \theta _{i},\theta
_{j}\right\} \right) \mathbf{,}
\label{eqn: pseudo expression of eta with uniform prior}
\end{eqnarray}%
where $\Theta _{i}=\left\{ 
\begin{array}{c}
\left[ a_{i},b_{i}-u_{i}\right] \text{ if }u_{i}>0, \\ 
\left[ a_{i}-u_{i},b_{i}\right] \text{ if }u_{i}<0,%
\end{array}%
\right. $ and $\Theta _{j}=\left\{ 
\begin{array}{c}
\left[ a_{j},b_{j}-v_{j}\right] \text{ if }v_{j}>0, \\ 
\left[ a_{j}-v_{j},b_{j}\right] \text{ if }v_{j}<0,%
\end{array}%
\right. $. For $i=j,$ one can have\textbf{\ $\mathbf{v}=\pm \mathbf{u}$, }%
then\textbf{\ }one obtains 
\begin{eqnarray}
\eta \left( \alpha ,\beta ,\mathbf{u},\mathbf{v}=\pm \mathbf{u}\right) 
&=&\int_{\Theta }\acute{\eta}_{\mathbf{\theta }}\left( \alpha ,\beta ,%
\mathbf{u},\mathbf{v}\right) \frac{p^{\alpha }\left( \theta
_{i}+u_{i}\right) p^{\beta }\left( \theta _{i}\pm u_{i}\right) }{p^{\alpha
+\beta -1}\left( \theta _{i}\right) }\overset{q}{\underset{k\neq i}{\underset%
{k=1}{\dprod }}}p\left( \theta _{k}\right) d\mathbf{\theta }  \notag \\
&=&\frac{1}{\overset{q}{\underset{k=1}{\dprod }}\left( b_{k}-a_{k}\right) }%
\int_{\Theta ^{q-1}}\int_{\Theta _{i}}\acute{\eta}_{\mathbf{\theta }}\left(
\alpha ,\beta ,\mathbf{u},\mathbf{v}=\pm \mathbf{u}\right) d\theta
_{i}d\left( \mathbf{\theta /}\left\{ \theta _{i}\right\} \right) \mathbf{.}
\label{eqn: pseudo expression of eta with uniform prior 2}
\end{eqnarray}

In the last equation, if\textbf{\ $\mathbf{v}=-\mathbf{u,}$ }then\ $\Theta
_{i}=\left\{ 
\begin{array}{c}
\left[ a_{i}+u_{i},b_{i}-u_{i}\right] \text{ if }u_{i}>0, \\ 
\left[ a_{i}-u_{i},b_{i}+u_{i}\right] \text{ if }u_{i}<0,%
\end{array}%
\right. ,$ while, if \textbf{$\mathbf{v}=\mathbf{u,}$} then $\Theta
_{i}=\left\{ 
\begin{array}{c}
\left[ a_{i},b_{i}-u_{i}\right] \text{ if }u_{i}>0, \\ 
\left[ a_{i}-u_{i},b_{i}\right] \text{ if }u_{i}<0,%
\end{array}%
.\right. $

Depending on the structure of $\acute{\eta}_{\mathbf{\theta }}\left( \alpha
,\beta ,\mathbf{u},\mathbf{v}\right) ,$ $\eta \left( \alpha ,\beta ,\mathbf{u%
},\mathbf{v}\right) $ has to be computed numerically or a closed-form
expression can be found.

Another particular case which appears sometimes is when the function $\acute{%
\eta}_{\mathbf{\theta }}\left( \alpha ,\beta ,\mathbf{u},\mathbf{v}\right) $
does not depend on $\mathbf{\theta }$ (see, \cite{Ren07}\cite{VTB07}\cite%
{BSEV97}\cite{BEV96a}\cite{NVT94}\cite{Ath01}\cite{BE08}\cite{BEV96b} and
Section \ref{Sec: Specific application} of this paper). In this case, $%
\acute{\eta}_{\mathbf{\theta }}\left( \alpha ,\beta ,\mathbf{u},\mathbf{v}%
\right) $ is denoted $\acute{\eta}\left( \alpha ,\beta ,\mathbf{u},\mathbf{v}%
\right) $ and one obtains from Eqn. (\ref{eqn: pseudo expression of eta with
uniform prior})%
\begin{eqnarray}
\eta \left( \alpha ,\beta ,\mathbf{u},\mathbf{v}\right)  &=&\frac{\acute{\eta%
}\left( \alpha ,\beta ,\mathbf{u},\mathbf{v}\right) }{\overset{q}{\underset{%
k=1}{\dprod }}\left( b_{k}-a_{k}\right) }\left( \overset{q}{\underset{k\neq
i,k\neq j}{\underset{k=1}{\dprod }}}\int_{a_{k}}^{b_{k}}d\theta _{k}\right)
\int_{\Theta _{i}}d\theta _{i}\int_{\Theta _{j}}d\theta _{j}  \notag \\
&=&\frac{\left( b_{i}-a_{i}-\left\vert u_{i}\right\vert \right) \left(
b_{j}-a_{j}-\left\vert v_{j}\right\vert \right) }{\left( b_{i}-a_{i}\right)
\left( b_{j}-a_{j}\right) }\acute{\eta}\left( \alpha ,\beta ,\mathbf{u},%
\mathbf{v}\right) ,
\end{eqnarray}%
and from Eqn. (\ref{eqn: pseudo expression of eta with uniform prior 2})%
\begin{equation}
\eta \left( \alpha ,\beta ,\mathbf{u},\mathbf{v}=\mathbf{u}\right) =\frac{%
\left( b_{i}-a_{i}-\left\vert u_{i}\right\vert \right) }{\left(
b_{i}-a_{i}\right) }\acute{\eta}\left( \alpha ,\beta ,\mathbf{u},\mathbf{v}%
\right) ,
\end{equation}%
and 
\begin{equation}
\eta \left( \alpha ,\beta ,\mathbf{u},\mathbf{v}=-\mathbf{u}\right) =\frac{%
\left( b_{i}-a_{i}-2\left\vert u_{i}\right\vert \right) }{\left(
b_{i}-a_{i}\right) }\acute{\eta}\left( \alpha ,\beta ,\mathbf{u},\mathbf{v}%
\right) .
\end{equation}

\subsection{Analysis of $\protect\eta \left( \protect\alpha ,\protect\beta ,%
\mathbf{u},\mathbf{v}\right) $ with a Gaussian prior}

Finally, one can mention that if the prior is now assumed to be Gaussian, 
\emph{i.e.}, $\theta _{i}\sim \mathcal{N}\left( \mu _{i},\sigma
_{i}^{2}\right) \forall i$ and $\acute{\eta}_{\mathbf{\theta }}\left( \alpha
,\beta ,\mathbf{u},\mathbf{v}\right) $ does not depend on $\mathbf{\theta }$
one obtains after a straightforward calculation%
\begin{eqnarray}
\eta \left( \alpha ,\beta ,\mathbf{u},\mathbf{v}\right) &=&\acute{\eta}%
\left( \alpha ,\beta ,\mathbf{u},\mathbf{v}\right) \int_{%
\mathbb{R}
}\frac{p^{\alpha }\left( \theta _{i}+u_{i}\right) }{p^{\alpha -1}\left(
\theta _{i}\right) }d\theta _{i}\int_{%
\mathbb{R}
}\frac{p^{\beta }\left( \theta _{j}+v_{j}\right) }{p^{\beta -1}\left( \theta
_{j}\right) }d\theta _{j}  \notag \\
&=&\acute{\eta}\left( \alpha ,\beta ,\mathbf{u},\mathbf{v}\right) \exp
\left( -\frac{1}{2}\left( \frac{\alpha \left( 1-\alpha \right) u_{i}^{2}}{%
\sigma _{i}^{2}}+\frac{\beta \left( 1-\beta \right) v_{j}^{2}}{\sigma
_{j}^{2}}\right) \right) ,
\end{eqnarray}%
\begin{eqnarray}
\eta \left( \alpha ,\beta ,\mathbf{u},\mathbf{v}=\mathbf{u}\right) &=&\acute{%
\eta}\left( \alpha ,\beta ,\mathbf{u},\mathbf{v}\right) \int_{%
\mathbb{R}
}\frac{p^{\alpha +\beta }\left( \theta _{i}+u_{i}\right) }{p^{\alpha +\beta
-1}\left( \theta _{i}\right) }d\theta _{i}  \notag \\
&=&\acute{\eta}\left( \alpha ,\beta ,\mathbf{u},\mathbf{v}\right) \exp
\left( -\frac{\left( \alpha +\beta \right) \left( 1-\alpha -\beta \right)
u_{i}^{2}}{2\sigma _{i}^{2}}\right) ,
\end{eqnarray}%
and%
\begin{eqnarray}
\eta \left( \alpha ,\beta ,\mathbf{u},\mathbf{v}=-\mathbf{u}\right) &=&%
\acute{\eta}\left( \alpha ,\beta ,\mathbf{u},\mathbf{v}\right) \int_{%
\mathbb{R}
}\frac{p^{\alpha }\left( \theta _{i}+u_{i}\right) p^{\beta }\left( \theta
_{i}-u_{i}\right) }{p^{\alpha +\beta -1}\left( \theta _{i}\right) }d\theta
_{i}  \notag \\
&=&\acute{\eta}\left( \alpha ,\beta ,\mathbf{u},\mathbf{v}\right) \exp
\left( -\frac{\left( \alpha +\beta -\alpha ^{2}-\beta ^{2}+2\alpha \beta
\right) u_{i}^{2}}{2\sigma _{i}^{2}}\right) .
\end{eqnarray}

\section{Specific applications to array processing: DOA estimation\label%
{Sec: Specific application}}

We now consider the application of the Weiss-Weinstein bound in the
particular context of source localization. Indeed, until now, the structure
of the steering matrix $\mathbf{A}\left( \mathbf{\theta }\right) $ for a
particular problem has not been used in the proposed (semi) closed-form
expressions. Consequently, these previous results can be applied to a large
class of estimation problems such as far-field and near-field sources
localization, passive localization with polarized array of sensors, or radar
(known waveforms).

Here, we want to focus on the direction-of-arrival estimation of a single
source in the far-field area with narrow-band signal. In this case, the
steering matrix $\mathbf{A}\left( \mathbf{\theta }\right) $ becomes a
steering vector denoted $\mathbf{a}\left( \mathbf{\theta }\right) $ (except
for one preliminary result concerning the conditional model which will be
given whatever the number of sources in Section \ref{Sec: Specific
application Preliminary Conditional}). The structure of this vector will be
specified by the analysis of two kinds of array geometry: the non-uniform
linear array from which only one angle-of-arrival can be estimated ($\mathbf{%
\theta }$ becomes a scalar) and the arbitrary planar array from which both
azimuth and elevation can be estimated ($\mathbf{\theta }$ becomes a $%
2\times 1$ vector). In any cases, the array always consists of $M$
identical, omnidirectional sensors. Both model $\mathcal{M}_{1}$ and $%
\mathcal{M}_{2}$ will be considered and the noise will be assumed spatially
uncorrelated: $\mathbf{R}_{\mathbf{n}}=\sigma _{n}^{2}\mathbf{I}$. Since we
focus on the single source scenario, the variance of the source signal $%
s\left( t\right) $ is denoted $\sigma _{s}^{2}$ for the model $\mathcal{M}%
_{1}$.

The general structure of the $i^{th}$ element of the steering vector is as
follows 
\begin{equation}
\left\{ \mathbf{a}\left( \mathbf{\theta }\right) \right\} _{i}=\exp \left( j%
\frac{2\pi }{\lambda }\mathbf{r}_{i}^{T}\mathbf{\theta }\right) ,\text{ }%
i=1,\ldots ,M  \label{eqn: general element steering vector}
\end{equation}%
where $\mathbf{\theta }$ represents the parameter vector, where $\lambda $
denotes the wavelength, and where $\mathbf{r}_{i}$ denotes the coordinate of
the $i^{th}$ sensor position with respect to a given referential. In the
following, $\mathbf{r}_{i}$ will be a scalar or a $2\times 1$ vector
depending on the context (linear array or planar array).

\subsection{Preliminary results}

Since our analysis is now reduced to the single source case, we give here
some other closed-form expressions which will be useful when we will detail
the specific linear and planar arrays.

\subsubsection{Unconditional observation model $\mathcal{M}_{1}$}

In order to detail the set of functions $\acute{\eta}_{\theta }$ given by
Eqn. (\ref{eqn: set of eta prime function unconditional model}), one has to
find closed-form expressions of the determinant $\left\vert \mathbf{R}_{%
\mathbf{y}}(\mathbf{\theta }+\mathbf{u})\right\vert $ and of determinants
having the following structure: $\left\vert {m}_{1}\mathbf{R}_{\mathbf{y}%
}^{-1}(\mathbf{\theta }_{1})+{m}_{2}\mathbf{R}_{\mathbf{y}}^{-1}(\mathbf{%
\theta }_{2})\right\vert $ with $m_{1}+m_{2}=1$ or $\left\vert {m}_{1}%
\mathbf{R}_{\mathbf{y}}^{-1}(\mathbf{\theta }_{1})+{m}_{2}\mathbf{R}_{%
\mathbf{y}}^{-1}(\mathbf{\theta }_{2})+{m}_{3}\mathbf{R}_{\mathbf{y}}^{-1}(%
\mathbf{\theta }_{3})\right\vert $ with $m_{1}+m_{2}+m_{3}=1$. Under $%
\mathcal{M}_{1}$, the observation covariance matrix is now given by 
\begin{equation}
\mathbf{R}_{\mathbf{y}}(\mathbf{\theta })=\sigma _{s}^{2}\mathbf{a}(\mathbf{%
\theta })\mathbf{a}^{H}(\mathbf{\theta })+\sigma _{n}^{2}\mathbf{I}_{M}.
\end{equation}

Concerning the calculation of $\left\vert \mathbf{R}_{\mathbf{y}}(\mathbf{%
\theta }+\mathbf{u})\right\vert $, it is easy to find%
\begin{equation}
\left\vert \mathbf{R}_{\mathbf{y}}(\mathbf{\theta }+\mathbf{u})\right\vert
=\sigma _{n}^{2M}\left( 1+\frac{\sigma _{s}^{2}}{\sigma _{n}^{2}}\left\Vert 
\mathbf{a(\theta +u)}\right\Vert ^{2}\right) .
\label{eqn: determinant of covariance matrix with 1 elements}
\end{equation}

Moreover, after calculation detailed in Appendix \ref{Sec: Appendix C}, one
obtains for the other determinants%
\begin{eqnarray}
\left\vert {m}_{1}\mathbf{R}_{\mathbf{y}}^{-1}(\mathbf{\theta }_{1})+{m}_{2}%
\mathbf{R}_{\mathbf{y}}^{-1}(\mathbf{\theta }_{2})\right\vert &=&\frac{1}{%
\left( \sigma _{n}^{2}\right) ^{M}}\left( 1-\varphi _{1}{m}_{1}\left\Vert 
\mathbf{a}(\mathbf{\theta }_{1})\right\Vert ^{2}+{m}_{2}\varphi
_{2}\left\Vert \mathbf{a}(\mathbf{\theta }_{2})\right\Vert ^{2}\right. 
\notag \\
&&\left. -\varphi _{1}{m}_{1}\varphi _{2}{m}_{2}\left( \left\Vert \mathbf{a}%
^{H}(\mathbf{\theta }_{1})\mathbf{a}(\mathbf{\theta }_{2})\right\Vert
^{2}-\left\Vert \mathbf{a}(\mathbf{\theta }_{1})\right\Vert ^{2}\left\Vert 
\mathbf{a}(\mathbf{\theta }_{2})\right\Vert ^{2}\right) \right)
\label{eqn: determinant of covariance matrix with 2 elements}
\end{eqnarray}%
and%
\begin{eqnarray}
&&\left\vert m_{1}\mathbf{R}_{\mathbf{y}}^{-1}(\mathbf{\theta }_{1})+m_{2}%
\mathbf{R}_{\mathbf{y}}^{-1}(\mathbf{\theta }_{2})+m_{3}\mathbf{R}_{\mathbf{y%
}}^{-1}(\mathbf{\theta }_{3})\right\vert =  \notag \\
&&\frac{1}{\left( \sigma _{n}^{2}\right) ^{M}}\left( 1-\underset{k=1}{%
\overset{3}{\sum }}{m}_{k}\varphi _{k}\left\Vert \mathbf{a}(\mathbf{\theta }%
_{k})\right\Vert ^{2}\right. -\frac{1}{2}\underset{k=1}{\overset{3}{\sum }}%
\underset{\underset{k^{\prime }\neq k}{k^{\prime }=1}}{\overset{3}{\sum }}%
m_{k}\varphi _{k}m_{k^{\prime }}\varphi _{k^{\prime }}\left( \left\Vert 
\mathbf{a}^{H}(\mathbf{\theta }_{k})\mathbf{a}(\mathbf{\theta }_{k^{\prime
}})\right\Vert ^{2}-\left\Vert \mathbf{a}(\mathbf{\theta }_{k})\right\Vert
^{2}\left\Vert \mathbf{a}(\mathbf{\theta }_{k^{\prime }})\right\Vert
^{2}\right)  \notag \\
&&-\left( \underset{k=1}{\overset{3}{\dprod }}{m}_{k}\varphi _{k}\right)
\left( \underset{k=1}{\overset{3}{\dprod }}\left\Vert \mathbf{a}(\mathbf{%
\theta }_{k})\right\Vert ^{2}-\frac{1}{2}\underset{k=1}{\overset{3}{\sum }}%
\underset{\underset{k^{\prime }\neq k}{k^{\prime }=1}}{\overset{3}{\sum }}%
\underset{\underset{k^{\prime \prime }\neq k^{\prime }\neq k}{k^{\prime
\prime }=1}}{\overset{3}{\sum }}\left\Vert \mathbf{a}^{H}(\mathbf{\theta }%
_{k})\mathbf{a}(\mathbf{\theta }_{k^{\prime }})\right\Vert ^{2}\left\Vert 
\mathbf{a}(\mathbf{\theta }_{k^{\prime \prime }})\right\Vert ^{2}\right. 
\notag \\
&&\left. \left. +\mathbf{a}^{H}(\mathbf{\theta }_{3})\mathbf{a}(\mathbf{%
\theta }_{2})\mathbf{a}^{H}(\mathbf{\theta }_{1})\mathbf{a}(\mathbf{\theta }%
_{3})\mathbf{a}^{H}(\mathbf{\theta }_{2})\mathbf{a}(\mathbf{\theta }_{1})+%
\mathbf{a}^{H}(\mathbf{\theta }_{3})\mathbf{a}(\mathbf{\theta }_{1})\mathbf{a%
}^{H}(\mathbf{\theta }_{1})\mathbf{a}(\mathbf{\theta }_{2})\mathbf{a}^{H}(%
\mathbf{\theta }_{2})\mathbf{a}(\mathbf{\theta }_{3})\right) \right) ,
\label{eqn: determinant of covariance matrix with 3 elements}
\end{eqnarray}%
where 
\begin{equation}
\varphi _{k}=\frac{\sigma _{s}^{2}}{\sigma _{s}^{2}\left\Vert \mathbf{a}(%
\mathbf{\theta }_{k})\right\Vert ^{2}+\sigma _{n}^{2}},\quad k=1,2,3.
\end{equation}

\subsubsection{Conditional observation model $\mathcal{M}_{2}$\label{Sec:
Specific application Preliminary Conditional}}

Note that the results proposed here are in the context of any number of
sources. Under the conditional model, the set of functions $\acute{\eta}%
_{\theta }$ given by Eqn. (\ref{eqn: set of eta prime function conditional
model}) is linked to the function $\zeta _{\mathbf{\theta }}\left( \mathbf{%
\mu ,\rho }\right) $ given by Eqn. (\ref{eqn: general function zeta}). In
this analysis, the vector $\mathbf{\mu }$ takes the value $\mu _{i}$ at the $%
i^{th}$ row and zero elsewhere and the vector $\mathbf{\rho }$ takes the
value $\rho _{j}$ at the $j^{th}$ row and zero elsewhere (of course, one can
has $i=j$). In Appendix \ref{Sec: Appendix D}, the calculation of the
following closed-form expressions for $\zeta _{\mathbf{\theta }}\left( 
\mathbf{\mu ,\rho }\right) $ are detailed.

\begin{itemize}
\item If $\left( m-1\right) p+1\leq i,j\leq mp,$ where $p$ denotes the
number of parameters per source, then, we have%
\begin{eqnarray}
\zeta _{\mathbf{\theta }}\left( \mathbf{\mu ,\rho }\right) &=&\overset{T}{%
\underset{t=1}{\sum }}\left\Vert \left\{ \mathbf{s}\left( t\right) \right\}
_{m}\right\Vert ^{2}\underset{i=1}{\overset{M}{\sum }}\underset{j=1}{\overset%
{M}{\sum }}\left\{ \mathbf{R}_{\mathbf{n}}^{-1}\right\} _{i,j}\exp \left( j%
\frac{2\pi }{\lambda }\left( \mathbf{r}_{j}^{T}-\mathbf{r}_{i}^{T}\right) 
\mathbf{\theta }_{m}\right)  \notag \\
&&\times \left( \exp \left( -j\frac{2\pi }{\lambda }\mathbf{r}_{i}^{T}%
\mathbf{\mu }_{m}\right) -\exp \left( -j\frac{2\pi }{\lambda }\mathbf{r}%
_{i}^{T}\mathbf{\rho }_{m}\right) \right) \left( \exp \left( j\frac{2\pi }{%
\lambda }\mathbf{r}_{j}^{T}\mathbf{\mu }_{m}\right) -\exp \left( j\frac{2\pi 
}{\lambda }\mathbf{r}_{j}^{T}\mathbf{\rho }_{m}\right) \right) .
\end{eqnarray}

\item Otherwise, if $\left( m-1\right) p+1\leq i\leq mp$ and $\left(
n-1\right) p+1\leq j\leq np$, then we have%
\begin{eqnarray}
&&\zeta _{\mathbf{\theta }}\left( \mathbf{\mu ,\rho }\right) =  \notag \\
&&\overset{T}{\underset{t=1}{\sum }}\left\Vert \left\{ \mathbf{s}\left(
t\right) \right\} _{m}\right\Vert ^{2}\underset{i=1}{\overset{M}{\sum }}%
\underset{j=1}{\overset{M}{\sum }}\left\{ \mathbf{R}_{\mathbf{n}%
}^{-1}\right\} _{i,j}\exp \left( j\frac{2\pi }{\lambda }\left( \mathbf{r}%
_{j}^{T}-\mathbf{r}_{i}^{T}\right) \mathbf{\theta }_{m}\right) \exp \left( -j%
\frac{2\pi }{\lambda }\mathbf{r}_{i}^{T}\mathbf{\mu }_{m}\right) \exp \left(
j\frac{2\pi }{\lambda }\mathbf{r}_{j}^{T}\mathbf{\mu }_{m}\right)  \notag \\
&&+\overset{T}{\underset{t=1}{\sum }}\left\Vert \left\{ \mathbf{s}\left(
t\right) \right\} _{n}\right\Vert ^{2}\underset{i=1}{\overset{M}{\sum }}%
\underset{j=1}{\overset{M}{\sum }}\left\{ \mathbf{R}_{\mathbf{n}%
}^{-1}\right\} _{i,j}\exp \left( j\frac{2\pi }{\lambda }\left( \mathbf{r}%
_{j}^{T}-\mathbf{r}_{i}^{T}\right) \mathbf{\theta }_{n}\right) \exp \left( -j%
\frac{2\pi }{\lambda }\mathbf{r}_{i}^{T}\mathbf{\rho }_{n}\right) \exp
\left( j\frac{2\pi }{\lambda }\mathbf{r}_{j}^{T}\mathbf{\rho }_{n}\right) 
\notag \\
&&-2\func{Re}\left( \overset{T}{\underset{t=1}{\sum }}\left\{ \mathbf{s}%
\left( t\right) \right\} _{m}^{\ast }\left\{ \mathbf{s}\left( t\right)
\right\} _{n}\right.  \notag \\
&&\qquad \times \left. \underset{i=1}{\overset{M}{\sum }}\underset{j=1}{%
\overset{M}{\sum }}\left\{ \mathbf{R}_{\mathbf{n}}^{-1}\right\} _{i,j}\exp
\left( j\frac{2\pi }{\lambda }\left( \mathbf{r}_{j}^{T}\mathbf{\theta }_{n}-%
\mathbf{r}_{i}^{T}\mathbf{\theta }_{m}\right) \right) \exp \left( -j\frac{%
2\pi }{\lambda }\mathbf{r}_{i}^{T}\mathbf{\mu }_{m}\right) \exp \left( j%
\frac{2\pi }{\lambda }\mathbf{r}_{j}^{T}\mathbf{\rho }_{n}\right) \right) .
\end{eqnarray}
\end{itemize}

In particular, if one assumes $\mathbf{R}_{n}=\sigma _{n}^{2}\mathbf{I}$,
then, several simplifications can be done:

\begin{itemize}
\item If $\left( m-1\right) p+1\leq i,j\leq mp,$ then%
\begin{equation}
\zeta _{\mathbf{\theta }}\left( \mathbf{\mu ,\rho }\right) =\frac{1}{\sigma
_{n}^{2}}\underset{i=1}{\overset{M}{\sum }}\left\Vert \exp \left( -j\frac{%
2\pi }{\lambda }\mathbf{r}_{i}^{T}\mathbf{\mu }_{m}\right) -\exp \left( -j%
\frac{2\pi }{\lambda }\mathbf{r}_{i}^{T}\mathbf{\rho }_{m}\right)
\right\Vert ^{2}\overset{T}{\underset{t=1}{\sum }}\left\Vert \left\{ \mathbf{%
s}\left( t\right) \right\} _{m}\right\Vert ^{2},
\end{equation}
where we note that the function $\zeta _{\mathbf{\theta }}\left( \mathbf{\mu
,\rho }\right) $ does not depend on the parameter $\mathbf{\theta }$.

\item Otherwise, if $\left( m-1\right) p+1\leq i\leq mp$ and $\left(
n-1\right) p+1\leq j\leq np$, then%
\begin{eqnarray}
\zeta _{\mathbf{\theta }}\left( \mathbf{\mu ,\rho }\right) &=&\frac{1}{%
\sigma _{n}^{2}}\underset{i=1}{\overset{M}{\sum }}\left\Vert \exp \left( -j%
\frac{2\pi }{\lambda }\mathbf{r}_{i}^{T}\mathbf{\mu }_{m}\right) \right\Vert
^{2}\overset{T}{\underset{t=1}{\sum }}\left\Vert \left\{ \mathbf{s}\left(
t\right) \right\} _{m}\right\Vert ^{2}+\frac{1}{\sigma _{n}^{2}}\underset{i=1%
}{\overset{M}{\sum }}\left\Vert \exp \left( -j\frac{2\pi }{\lambda }\mathbf{r%
}_{i}^{T}\mathbf{\rho }_{n}\right) \right\Vert ^{2}\overset{T}{\underset{t=1}%
{\sum }}\left\Vert \left\{ \mathbf{s}\left( t\right) \right\}
_{n}\right\Vert ^{2}  \notag \\
&&\hspace{-2cm}-2\func{Re}\left( \frac{1}{\sigma _{n}^{2}}\underset{i=1}{%
\overset{M}{\sum }}\exp \left( j\frac{2\pi }{\lambda }\mathbf{r}%
_{i}^{T}\left( \mathbf{\theta }_{n}-\mathbf{\theta }_{m}\right) \right) \exp
\left( -j\frac{2\pi }{\lambda }\mathbf{r}_{i}^{T}\mathbf{\mu }_{m}\right)
\exp \left( j\frac{2\pi }{\lambda }\mathbf{r}_{i}^{T}\mathbf{\rho }%
_{n}\right) \overset{T}{\underset{t=1}{\sum }}\left\{ \mathbf{s}\left(
t\right) \right\} _{m}^{\ast }\left\{ \mathbf{s}\left( t\right) \right\}
_{n}\right)
\end{eqnarray}
\end{itemize}

It is clear that the proposed above formulas for both the unconditional and
the conditional models can be applied to any kind of array geometry and
whatever the number of sources. However, they generally depend on the
parameter vector $\mathbf{\theta }$. This means that, in general, the
calculation of the set of functions $\eta $ will have to be performed
numerically (except if one is able to find a closed-form expression of Eqn. (%
\ref{eqn: general function eta rewritten})). In the following we present a
kind of array geometry where, fortunately, the set of functions $\acute{\eta}%
_{\theta }$ will not depend on $\mathbf{\theta }$ leading to a
straightforward calculation of the bound.

\subsection{3D Source localization with a planar array}

\begin{figure}[t]
\begin{center}
\epsfig{figure=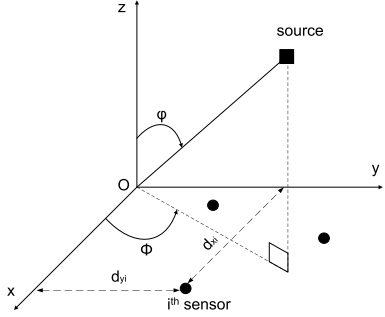,width=50mm}
\end{center}
\caption{ 3D source localization using a planar array antenna.}
\label{fig:Coordinate}
\end{figure}

We first consider the problem of DOA estimation of a single narrow band
source in the far field area by using an arbitrary planar array. In fact, we
start by this general setting because the non-uniform linear array is
clearly a particular case of this array. Without loss of generality, we
assume that the sensors of this array lay on the $xOy$ plan with Cartesian
coordinates (see Fig. \ref{fig:Coordinate}). Therefore, the vector $\mathbf{r%
}_{i}$ contains the coordinate of the $i^{th}$ sensor position with respect
to this referential, \emph{i.e.}, $\mathbf{r}_{i}=\left[ d_{x_{i}}\ d_{y_{i}}%
\right] ^{T},$ $i=1,\ldots ,M$. From (\ref{eqn: general element steering
vector}), the steering vector is given by 
\begin{equation}
\mathbf{a(\theta )}=\left[ \exp {\left( j\frac{2\pi }{\lambda }%
(d_{x_{1}}u+d_{y_{_{1}}}v)\right) \ldots }\exp {\left( j\frac{2\pi }{\lambda 
}(d_{x_{M}}u+d_{y_{M}}v)\right) }\right] ^{T},
\end{equation}%
where, as in \cite{BEV96a}, the parameter vector of interest is $\mathbf{%
\theta }=[u\quad v]^{T}$ where 
\begin{equation}
\left\{ 
\begin{array}{l}
u=\sin {\varphi }\cos {\phi ,} \\ 
v=\sin {\varphi }\sin {\phi ,}%
\end{array}%
\right. 
\end{equation}%
and where $\varphi $ and $\phi $ represent the elevation and azimuth angles
of the source, respectively. The parameters space is such that $u\in \left[
-1,1\right] $ and $v\in \left[ -1,1\right] $. Therefore, we assume that they
both follow a uniform distribution over $\left[ -1,1\right] $. Note that
from a physical point of view, it should be more tempting to choose a
uniform prior for $\varphi $ and $\phi $. This will lead to a probability
density functions for $u$ and $v$ not uniform. To the best of our knowledge,
this assumption has only been used in the context of lower bounds in \cite%
{NVT94}. Unfortunately, such prior leads to an untractable expression of the
bound (see Eqn. (21) of \cite{NVT94}). Consequently, other authors have
generally not specified the prior leading to semi closed-form expressions of
bounds (\emph{i.e. }that it remains a numerical integration to perform over
the parameters) \cite{NVT94}\cite{Bel95}\cite{XBR04}. On the other hand, in
order to obtain a closed-form expression, authors have generally used a
simplified assumption, \emph{i.e.} a uniform prior directly on $u$ and $v$
(see, for example, \cite{Ath01}\cite{XBB04})\textbf{. }In this paper, we
have followed the same way by expecting a slight modification of performance
with respect to a more physical model and in order to be able to get
closed-form expressions of the bound.

We choose the matrix of test points such that 
\begin{equation}
\mathbf{H}=[\mathbf{h}_{u}\quad \mathbf{h}_{v}]=\left[ 
\begin{array}{cc}
h_{u} & 0 \\ 
0 & h_{v}%
\end{array}%
\right] .
\end{equation}

Then, we have: $\mathbf{\theta }\mathbf{+h_{u}}=[u+h_{u}\quad v]^{T}$ and $%
\mathbf{\theta }\mathbf{+h_{v}}=[u\quad v+h_{v}]^{T}$. Moreover, we now have
two elements $s_{i}\in \left[ 0,1\right] ,$ $i=1,2$ for which we will prefer
the notation $s_{u}$ and $s_{v},$ respectively.

\subsubsection{Unconditional observation model $\mathcal{M}_{1}$}

Under $\mathcal{M}_{1}$, let us set $U_{SNR}=\frac{\sigma _{s}^{4}}{\sigma
_{n}^{2}(M\sigma _{s}^{2}+\sigma _{n}^{2})}$. The closed-form expressions of
the elements of matrix $\mathbf{G=}\left[ 
\begin{array}{cc}
\{\mathbf{G}\}_{uu} & \{\mathbf{G}\}_{uv} \\ 
\{\mathbf{G}\}_{vu} & \{\mathbf{G}\}_{vv}%
\end{array}%
\right] $ are given by (see Appendix \ref{Sec: Appendix E} for the proof): 
\begin{equation}
\{\mathbf{G}\}_{uu}=\frac{{\small \left( 
\begin{array}{l}
\left( 1-\frac{|h_{u}|}{2}\right) \left( 1+2s_{u}(1-2s_{u})U_{SNR}\left(
M^{2}-\left\Vert \overset{M}{\underset{k=1}{\sum }}\exp {\left( -j\frac{2\pi 
}{\lambda }d_{x_{k}}h_{u}\right) }\right\Vert ^{2}\right) \right) ^{-T} \\ 
+\left( 1-\frac{|h_{u}|}{2}\right) \left(
1+2(1-s_{u})(2s_{u}-1)U_{SNR}\left( M^{2}-\left\Vert \overset{M}{\underset{%
k=1}{\sum }}\exp {\left( -j\frac{2\pi }{\lambda }d_{x_{k}}h_{u}\right) }%
\right\Vert ^{2}\right) \right) ^{-T} \\ 
-2\left( 1-|h_{u}|\right) \left( 1+s_{u}(1-s_{u})U_{SNR}\left(
M^{2}-\left\Vert \overset{M}{\underset{k=1}{\sum }}\exp {\left( -j\frac{4\pi 
}{\lambda }d_{x_{k}}h_{u}\right) }\right\Vert ^{2}\right) \right) ^{-T}%
\end{array}%
\right) }}{\left( 1-\frac{|h_{u}|}{2}\right) ^{2}\left(
1+s_{u}(1-s_{u})U_{SNR}\left( M^{2}-\left\Vert \overset{M}{\underset{k=1}{%
\sum }}\exp {\left( -j\frac{2\pi }{\lambda }d_{x_{k}}h_{u}\right) }%
\right\Vert ^{2}\right) \right) ^{-2T}},  \label{eqn: UWWB planar Guu}
\end{equation}%
\begin{equation}
\{\mathbf{G}\}_{vv}=\frac{{\small \left( 
\begin{array}{l}
\left( 1-\frac{|h_{v}|}{2}\right) \left( 1+2s_{v}(1-2s_{v})U_{SNR}\left(
M^{2}-\left\Vert \overset{M}{\underset{k=1}{\sum }}\exp {\left( -j\frac{2\pi 
}{\lambda }d_{y_{k}}h_{v}\right) }\right\Vert ^{2}\right) \right) ^{-T} \\ 
+\left( 1-\frac{|h_{v}|}{2}\right) \left(
1+2(1-s_{v})(2s_{v}-1)U_{SNR}\left( M^{2}-\left\Vert \overset{M}{\underset{%
k=1}{\sum }}\exp {\left( -j\frac{2\pi }{\lambda }d_{y_{k}}h_{v}\right) }%
\right\Vert ^{2}\right) \right) ^{-T} \\ 
-2\left( 1-|h_{v}|\right) \left( 1+s_{v}(1-s_{v})U_{SNR}\left(
M^{2}-\left\Vert \overset{M}{\underset{k=1}{\sum }}\exp {\left( -j\frac{4\pi 
}{\lambda }d_{y_{k}}h_{v}\right) }\right\Vert ^{2}\right) \right) ^{-T}%
\end{array}%
\right) }}{\left( 1-\frac{|h_{v}|}{2}\right) ^{2}\left(
1+s_{v}(1-s_{v})U_{SNR}\left( M^{2}-\left\Vert \overset{M}{\underset{k=1}{%
\sum }}\exp {\left( -j\frac{2\pi }{\lambda }d_{y_{k}}h_{v}\right) }%
\right\Vert ^{2}\right) \right) ^{-2T}},  \label{eqn: UWWB planar Gvv}
\end{equation}%
\begin{equation}
{\scriptsize \{\mathbf{G}\}_{uv}=\frac{\left( 
\begin{array}{l}
\left( 
\begin{array}{l}
1-U_{SNR}\left( 
\begin{array}{l}
s_{u}s_{v}\left( \left\Vert \overset{M}{\underset{k=1}{\sum }}\exp {\left( -j%
\frac{2\pi }{\lambda }(d_{x_{k}}h_{u}-d_{y_{k}}h_{v})\right) }\right\Vert
^{2}-M^{2}\right) \\ 
+s_{u}(1-s_{u}-s_{v})\left( \left\Vert \overset{M}{\underset{k=1}{\sum }}%
\exp {\left( -j\frac{2\pi }{\lambda }d_{x_{k}}h_{u}\right) }\right\Vert
^{2}-M^{2}\right) \\ 
+s_{v}(1-s_{u}-s_{v})\left( \left\Vert \overset{M}{\underset{k=1}{\sum }}%
\exp {\left( -j\frac{2\pi }{\lambda }d_{y_{k}}h_{v}\right) }\right\Vert
^{2}-M^{2}\right)%
\end{array}%
\right) \\ 
-s_{u}s_{v}(1-s_{u}-s_{v})\frac{U_{SNR}^{2}\sigma _{n}^{2}}{\sigma _{s}^{2}}%
\times \\ 
\times \left( 
\begin{array}{l}
\overset{M}{\underset{k=1}{\sum }}\exp {\left( j\frac{2{\pi }d_{y_{k}}h_{v}}{%
\lambda }\right) }\overset{M}{\underset{k=1}{\sum }}\exp {\left( -j\frac{2{%
\pi }d_{x_{k}}h_{u}}{\lambda }\right) }\overset{M}{\underset{k=1}{\sum }}%
\exp {\left( j\frac{2\pi (d_{x_{k}}h_{u}-d_{y_{k}}h_{v})}{\lambda }\right) }
\\ 
+\overset{M}{\underset{k=1}{\sum }}\exp {\left( -j\frac{2{\pi }d_{y_{k}}h_{v}%
}{\lambda }\right) }\overset{M}{\underset{k=1}{\sum }}\exp {\left( j\frac{2{%
\pi }d_{x_{k}}h_{u}}{\lambda }\right) }\overset{M}{\underset{k=1}{\sum }}%
\exp {\left( -j\frac{2\pi (d_{x_{k}}h_{u}-d_{y_{k}}h_{v})}{\lambda }\right) }
\\ 
-M\left\Vert \overset{M}{\underset{k=1}{\sum }}\exp {\left( -j\frac{2\pi }{%
\lambda }d_{y_{k}}h_{v}\right) }\right\Vert ^{2}-M\left\Vert \overset{M}{%
\underset{k=1}{\sum }}\exp {\left( -j\frac{2\pi }{\lambda }%
d_{x_{k}}h_{u}\right) }\right\Vert ^{2} \\ 
-M\left\Vert \overset{M}{\underset{k=1}{\sum }}\exp {\left( -j\frac{2\pi }{%
\lambda }(d_{x_{k}}h_{u}-d_{y_{k}}h_{v})\right) }\right\Vert ^{2}+M^{3}%
\end{array}%
\right)%
\end{array}%
\right) ^{-T} \\ 
+\left( 
\begin{array}{l}
1-U_{SNR}\left( 
\begin{array}{l}
(1-s_{u})(1-s_{v})\left( \left\Vert \overset{M}{\underset{k=1}{\sum }}\exp {%
\left( j\frac{2\pi }{\lambda }(d_{x_{k}}h_{u}-d_{y_{k}}h_{v})\right) }%
\right\Vert ^{2}-M^{2}\right) \\ 
+(1-s_{u})(s_{u}+s_{v}-1)\left( \left\Vert \overset{M}{\underset{k=1}{\sum }}%
\exp {\left( j\frac{2\pi }{\lambda }d_{x_{k}}h_{u}\right) }\right\Vert
^{2}-M^{2}\right) \\ 
+(1-s_{v})(s_{u}+s_{v}-1)\left( \left\Vert \overset{M}{\underset{k=1}{\sum }}%
\exp {\left( j\frac{2\pi }{\lambda }d_{y_{k}}h_{v}\right) }\right\Vert
^{2}-M^{2}\right)%
\end{array}%
\right) \\ 
-(1-s_{u})(1-s_{v})(s_{u}+s_{v}-1)\frac{U_{SNR}^{2}\sigma _{n}^{2}}{\sigma
_{s}^{2}}\times \\ 
\times \left( 
\begin{array}{l}
\overset{M}{\underset{k=1}{\sum }}\exp {\left( j\frac{2{\pi }d_{y_{k}}h_{v}}{%
\lambda }\right) }\overset{M}{\underset{k=1}{\sum }}\exp {\left( -j\frac{2{%
\pi }d_{x_{k}}h_{u}}{\lambda }\right) }\overset{M}{\underset{k=1}{\sum }}%
\exp {\left( j\frac{2\pi (d_{x_{k}}h_{u}-d_{y_{k}}h_{v})}{\lambda }\right) }
\\ 
+\overset{M}{\underset{k=1}{\sum }}\exp {\left( -j\frac{2{\pi }d_{y_{k}}h_{v}%
}{\lambda }\right) }\overset{M}{\underset{k=1}{\sum }}\exp {\left( j\frac{2{%
\pi }d_{x_{k}}h_{u}}{\lambda }\right) }\overset{M}{\underset{k=1}{\sum }}%
\exp {\left( -j\frac{2\pi (d_{x_{k}}h_{u}-d_{y_{k}}h_{v})}{\lambda }\right) }
\\ 
-M\left\Vert \overset{M}{\underset{k=1}{\sum }}\exp {\left( -j\frac{2\pi }{%
\lambda }d_{y_{k}}h_{v}\right) }\right\Vert ^{2}-M\left\Vert \overset{M}{%
\underset{k=1}{\sum }}\exp {\left( -j\frac{2\pi }{\lambda }%
d_{x_{k}}h_{u}\right) }\right\Vert ^{2} \\ 
-M\left\Vert \overset{M}{\underset{k=1}{\sum }}\exp {\left( -j\frac{2\pi }{%
\lambda }(d_{x_{k}}h_{u}-d_{y_{k}}h_{v})\right) }\right\Vert ^{2}+M^{3}%
\end{array}%
\right)%
\end{array}%
\right) ^{-T} \\ 
-\left( 
\begin{array}{l}
1-U_{SNR}\left( 
\begin{array}{l}
s_{u}(1-s_{v})\left( \left\Vert \overset{M}{\underset{k=1}{\sum }}\exp {%
\left( -j\frac{2\pi }{\lambda }(d_{x_{k}}h_{u}+d_{y_{k}}h_{v})\right) }%
\right\Vert ^{2}-M^{2}\right) \\ 
+s_{u}(s_{v}-s_{u})\left( \left\Vert \overset{M}{\underset{k=1}{\sum }}\exp {%
\left( -j\frac{2\pi }{\lambda }d_{x_{k}}h_{u}\right) }\right\Vert
^{2}-M^{2}\right) \\ 
+(1-s_{v})(s_{v}-s_{u})\left( \left\Vert \overset{M}{\underset{k=1}{\sum }}%
\exp {\left( j\frac{2\pi }{\lambda }d_{y_{k}}h_{v}\right) }\right\Vert
^{2}-M^{2}\right)%
\end{array}%
\right) \\ 
-s_{u}(1-s_{v})(s_{v}-s_{u})\frac{U_{SNR}^{2}\sigma _{n}^{2}}{\sigma _{s}^{2}%
}\times \\ 
\times \left( 
\begin{array}{l}
\overset{M}{\underset{k=1}{\sum }}\exp {\left( j\frac{2{\pi }d_{y_{k}}h_{v}}{%
\lambda }\right) }\overset{M}{\underset{k=1}{\sum }}\exp {\left( j\frac{2{%
\pi }d_{x_{k}}h_{u}}{\lambda }\right) }\overset{M}{\underset{k=1}{\sum }}%
\exp {\left( -j\frac{2\pi (d_{x_{k}}h_{u}+d_{y_{k}}h_{v})}{\lambda }\right) }
\\ 
+\overset{M}{\underset{k=1}{\sum }}\exp {\left( -j\frac{2{\pi }d_{y_{k}}h_{v}%
}{\lambda }\right) }\overset{M}{\underset{k=1}{\sum }}\exp {\left( -j\frac{2{%
\pi }d_{x_{k}}h_{u}}{\lambda }\right) }\overset{M}{\underset{k=1}{\sum }}%
\exp {\left( j\frac{2\pi (d_{x_{k}}h_{u}+d_{y_{k}}h_{v})}{\lambda }\right) }
\\ 
-M\left\Vert \overset{M}{\underset{k=1}{\sum }}\exp {\left( j\frac{2\pi }{%
\lambda }d_{y_{k}}h_{v}\right) }\right\Vert ^{2}-M\left\Vert \overset{M}{%
\underset{k=1}{\sum }}\exp {\left( -j\frac{2\pi }{\lambda }%
d_{x_{k}}h_{u}\right) }\right\Vert ^{2} \\ 
-M\left\Vert \overset{M}{\underset{k=1}{\sum }}\exp {\left( -j\frac{2\pi }{%
\lambda }(d_{x_{k}}h_{u}+d_{y_{k}}h_{v})\right) }\right\Vert ^{2}+M^{3}%
\end{array}%
\right)%
\end{array}%
\right) ^{-T} \\ 
-\left( 
\begin{array}{l}
1-U_{SNR}\left( 
\begin{array}{l}
s_{v}(1-s_{u})\left( \left\Vert \overset{M}{\underset{k=1}{\sum }}\exp {%
\left( -j\frac{2\pi }{\lambda }(d_{x_{k}}h_{u}+d_{y_{k}}h_{v})\right) }%
\right\Vert ^{2}-M^{2}\right) \\ 
+s_{v}(s_{u}-s_{v})\left( \left\Vert \overset{M}{\underset{k=1}{\sum }}\exp {%
\left( -j\frac{2\pi }{\lambda }d_{x_{k}}h_{u}\right) }\right\Vert
^{2}-M^{2}\right) \\ 
+(1-s_{u})(s_{u}-s_{v})\left( \left\Vert \overset{M}{\underset{k=1}{\sum }}%
\exp {\left( -j\frac{2\pi }{\lambda }d_{y_{k}}h_{v}\right) }\right\Vert
^{2}-M^{2}\right)%
\end{array}%
\right) \\ 
-s_{v}(1-s_{u})(s_{u}-s_{v})\frac{U_{SNR}^{2}\sigma _{n}^{2}}{\sigma _{s}^{2}%
}\times \\ 
\times \left( 
\begin{array}{l}
\overset{M}{\underset{k=1}{\sum }}\exp {\left( j\frac{2{\pi }d_{y_{k}}h_{v}}{%
\lambda }\right) }\overset{M}{\underset{k=1}{\sum }}\exp {\left( j\frac{2{%
\pi }d_{x_{k}}h_{u}}{\lambda }\right) }\overset{M}{\underset{k=1}{\sum }}%
\exp {\left( -j\frac{2\pi (d_{x_{k}}h_{u}+d_{y_{k}}h_{v})}{\lambda }\right) }
\\ 
+\overset{M}{\underset{k=1}{\sum }}\exp {\left( -j\frac{2{\pi }d_{y_{k}}h_{v}%
}{\lambda }\right) }\overset{M}{\underset{k=1}{\sum }}\exp {\left( -j\frac{2{%
\pi }d_{x_{k}}h_{u}}{\lambda }\right) }\overset{M}{\underset{k=1}{\sum }}%
\exp {\left( j\frac{2\pi (d_{x_{k}}h_{u}+d_{y_{k}}h_{v})}{\lambda }\right) }
\\ 
-M\left\Vert \overset{M}{\underset{k=1}{\sum }}\exp {\left( -j\frac{2\pi }{%
\lambda }d_{y_{k}}h_{v}\right) }\right\Vert ^{2}-M\left\Vert \overset{M}{%
\underset{k=1}{\sum }}\exp {\left( -j\frac{2\pi }{\lambda }%
d_{x_{k}}h_{u}\right) }\right\Vert ^{2} \\ 
-M\left\Vert \overset{M}{\underset{k=1}{\sum }}\exp {\left( -j\frac{2\pi }{%
\lambda }(d_{x_{k}}h_{u}+d_{y_{k}}h_{v})\right) }\right\Vert ^{2}+M^{3}%
\end{array}%
\right)%
\end{array}%
\right) ^{-T}%
\end{array}%
\right) }{%
\begin{array}{l}
\left( 1+s_{u}(1-s_{u})U_{SNR}\left( M^{2}-\left\Vert \overset{M}{\underset{%
k=1}{\sum }}\exp {\left( -j\frac{2\pi }{\lambda }d_{x_{k}}h_{u}\right) }%
\right\Vert ^{2}\right) \right) ^{-T}\left( 1+s_{v}(1-s_{v})U_{SNR}\left(
M^{2}-\left\Vert \overset{M}{\underset{k=1}{\sum }}\exp {\left( -j\frac{2\pi 
}{\lambda }d_{y_{k}}h_{v}\right) }\right\Vert ^{2}\right) \right) ^{-T}%
\end{array}%
},}  \label{eqn: UWWB planar Guv}
\end{equation}%
and, of course, $\{\mathbf{G}\}_{uv}=\{\mathbf{G}\}_{vu}$. Consequently, the
unconditional Weiss-Weinstein bound is $2\times 2$ matrix given by: 
\begin{eqnarray}
\mathbf{UWWB} &=&\mathbf{H}\mathbf{G}^{-1}\mathbf{H}^{T}  \notag \\
&=&\frac{1}{\{\mathbf{G}\}_{uu}\{\mathbf{G}\}_{vv}-\{\mathbf{G}\}_{uv}^{2}}%
\left[ 
\begin{array}{cc}
h_{u}^{2}\{\mathbf{G}\}_{vv} & -h_{u}h_{v}\{\mathbf{G}\}_{uv} \\ 
-h_{u}h_{v}\{\mathbf{G}\}_{uv} & h_{v}^{2}\{\mathbf{G}\}_{uu}%
\end{array}%
\right] ,  \label{eqn: UWWB planar with s}
\end{eqnarray}%
which has to be optimized over $s_{u},s_{v},h_{u},$ and $h_{v}.$ Concerning
the optimization over $s_{u}$ and $s_{v},$ several other works in the
literature have suggested to simply use $s_{u}=s_{v}=1/2.$ Most of the time,
numerical simulations of this simplified bound compared with the bound
obtained after optimization over $s_{u}$ and $s_{v}$ leads to the same
results while their is no formal proof of this fact (see \cite{VTB07} page
41 footnote 17). Note that, thanks to the expressions obtained in the next
Section concerning the linear array, we will be able to prove that $s=1/2$
is a (maybe not unique) correct choice for any linear array. In the case of
the planar array treated in this Section, we will only check this property
by simulation.

In the particular case where $s_{u}=s_{v}=1/2$ one obtains the following
simplified expressions 
\begin{eqnarray}
\{\mathbf{G}\}_{uu} &=&\frac{2\left( 1-\frac{|h_{u}|}{2}\right) -2\left(
1-|h_{u}|\right) \left( 1+\frac{U_{SNR}}{4}\left( M^{2}-\left\Vert \overset{M%
}{\underset{k=1}{\sum }}\exp {\left( -j\frac{4\pi }{\lambda }%
d_{x_{k}}h_{u}\right) }\right\Vert ^{2}\right) \right) ^{-T}}{\left( 1-\frac{%
|h_{u}|}{2}\right) ^{2}\left( 1+\frac{U_{SNR}}{4}\left( M^{2}-\left\Vert 
\overset{M}{\underset{k=1}{\sum }}\exp {\left( -j\frac{2\pi }{\lambda }%
d_{x_{k}}h_{u}\right) }\right\Vert ^{2}\right) \right) ^{-2T}},
\label{eqn: CF WWB unconditional planar uu} \\
\{\mathbf{G}\}_{vv} &=&\frac{2\left( 1-\frac{|h_{v}|}{2}\right) -2\left(
1-|h_{v}|\right) \left( 1+\frac{U_{SNR}}{4}\left( M^{2}-\left\Vert \overset{M%
}{\underset{k=1}{\sum }}\exp {\left( -j\frac{4\pi }{\lambda }%
d_{y_{k}}h_{v}\right) }\right\Vert ^{2}\right) \right) ^{-T}}{\left( 1-\frac{%
|h_{v}|}{2}\right) ^{2}\left( 1+\frac{U_{SNR}}{4}\left( M^{2}-\left\Vert 
\overset{M}{\underset{k=1}{\sum }}\exp {\left( -j\frac{2\pi }{\lambda }%
d_{y_{k}}h_{v}\right) }\right\Vert ^{2}\right) \right) ^{-2T}},
\label{eqn: CF WWB unconditional planar vv}
\end{eqnarray}%
and%
\begin{equation}
\{\mathbf{G}\}_{uv}=\frac{\left( 
\begin{array}{l}
2\left( 1+\frac{U_{SNR}}{4}\left( M^{2}-\left\Vert \overset{M}{\underset{k=1}%
{\sum }}\exp {\left( -j\frac{2\pi }{\lambda }(d_{x_{k}}h_{u}-d_{y_{k}}h_{v})%
\right) }\right\Vert ^{2}\right) \right) ^{-T} \\ 
-2\left( 1+\frac{U_{SNR}}{4}\left( M^{2}-\left\Vert \overset{M}{\underset{k=1%
}{\sum }}\exp {\left( -j\frac{2\pi }{\lambda }%
(d_{x_{k}}h_{u}+d_{y_{k}}h_{v})\right) }\right\Vert ^{2}\right) \right) ^{-T}%
\end{array}%
\right) }{\left( 1+\frac{U_{SNR}}{4}\left( M^{2}-\left\Vert \overset{M}{%
\underset{k=1}{\sum }}\exp {\left( -j\frac{2\pi }{\lambda }%
d_{x_{k}}h_{u}\right) }\right\Vert ^{2}\right) \right) ^{-T}\left( 1+\frac{%
U_{SNR}}{4}\left( M^{2}-\left\Vert \overset{M}{\underset{k=1}{\sum }}\exp {%
\left( -j\frac{2\pi }{\lambda }d_{y_{k}}h_{v}\right) }\right\Vert
^{2}\right) \right) ^{-T}}.  \label{eqn: CF WWB unconditional planar uv}
\end{equation}

Again, the Weiss-Weinstein bound is obtained by using the above expressions
in Eqn. (\ref{eqn: UWWB planar with s}) and after an optimization over the
test points. The optimization over the test points can be done over a search
grid or by using the ambiguity diagram of the array in order to reduce
significantly the computational cost (see \cite{RM97},\cite{XBR04}, \cite%
{RM95},\cite{TK99},\cite{RAFL07}).

\subsubsection{Conditional observation model $\mathcal{M}_{2}$\label{Sec:
Specific application PA Conditional model}}

Under $\mathcal{M}_{2}$, let us set $C_{SNR}=\frac{1}{\sigma _{n}^{2}}%
\overset{T}{\underset{t=1}{\sum }}{\left\Vert s(t)\right\Vert ^{2}}.$ The
closed-form expressions of the elements of matrix $\mathbf{G}$ are given by
(see Appendix \ref{Sec: Appendix F} for the proof): 
\begin{equation}
\left\{ \mathbf{G}\right\} _{uu}=\frac{{\small \left( 
\begin{array}{l}
\left( 1-\frac{|h_{u}|}{2}\right) \exp {\left( 4s_{u}(2s_{u}-1)C_{SNR}\left(
M-\overset{M}{\underset{k=1}{\sum }}\cos {\left( \frac{2\pi }{\lambda }%
d_{x_{k}}h_{u}\right) }\right) \right) } \\ 
+\left( 1-\frac{|h_{u}|}{2}\right) \exp {\left(
4(2s_{u}-1)(s_{u}-1)C_{SNR}\left( M-\overset{M}{\underset{k=1}{\sum }}\cos {%
\left( \frac{2\pi }{\lambda }d_{x_{k}}h_{u}\right) }\right) \right) } \\ 
-2(1-|h_{u}|)\exp {\left( 2s_{u}(s_{u}-1)C_{SNR}\left( M-\overset{M}{%
\underset{k=1}{\sum }}\cos {\left( \frac{4\pi }{\lambda }d_{x_{k}}h_{u}%
\right) }\right) \right) }%
\end{array}%
\right) }}{\left( 1-\frac{|h_{u}|}{2}\right) ^{2}\exp {\left(
4s_{u}(s_{u}-1)C_{SNR}\left( M-\overset{M}{\underset{k=1}{\sum }}\cos {%
\left( \frac{2\pi }{\lambda }d_{x_{k}}h_{u}\right) }\right) \right) }},
\label{eqn: CWWB planar Guu}
\end{equation}%
\begin{equation}
\left\{ \mathbf{G}\right\} _{vv}=\frac{{\small \left( 
\begin{array}{l}
\left( 1-\frac{|h_{v}|}{2}\right) \exp {\left( 4s_{v}(2s_{v}-1)C_{SNR}\left(
M-\overset{M}{\underset{k=1}{\sum }}\cos {\left( \frac{2\pi }{\lambda }%
d_{y_{k}}h_{v}\right) }\right) \right) } \\ 
+\left( 1-\frac{|h_{v}|}{2}\right) \exp {\left(
4(2s_{v}-1)(s_{v}-1)C_{SNR}\left( M-\overset{M}{\underset{k=1}{\sum }}\cos {%
\left( \frac{2\pi }{\lambda }d_{y_{k}}h_{v}\right) }\right) \right) } \\ 
-2(1-|h_{v}|)\exp {\left( 2s_{v}(s_{v}-1)C_{SNR}\left( M-\overset{M}{%
\underset{k=1}{\sum }}\cos {\left( \frac{4\pi }{\lambda }d_{y_{k}}h_{v}%
\right) }\right) \right) }%
\end{array}%
\right) }}{\left( 1-\frac{|h_{v}|}{2}\right) ^{2}\exp {\left(
4s_{v}(s_{v}-1)C_{SNR}\left( M-\overset{M}{\underset{k=1}{\sum }}\cos {%
\left( \frac{2\pi }{\lambda }d_{y_{k}}h_{v}\right) }\right) \right) }},
\label{eqn: CWWB planar Gvv}
\end{equation}%
\begin{equation}
\left\{ \mathbf{G}\right\} _{uv}=\frac{\left( 
\begin{array}{l}
\exp {\left( {\scriptsize 
\begin{array}{l}
2s_{u}(s_{u}+s_{v}-1)C_{SNR}\left( M-\overset{M}{\underset{k=1}{\sum }}\cos {%
\left( \frac{2\pi }{\lambda }d_{x_{k}}h_{u}\right) }\right) \\ 
+2s_{v}(s_{u}+s_{v}-1)C_{SNR}\left( M-\overset{M}{\underset{k=1}{\sum }}\cos 
{\left( \frac{2\pi }{\lambda }d_{y_{k}}h_{v}\right) }\right) \\ 
-2s_{u}s_{v}C_{SNR}\left( M-\overset{M}{\underset{k=1}{\sum }}\cos {\left( 
\frac{2\pi }{\lambda }(d_{x_{k}}h_{u}-d_{y_{k}}h_{v})\right) }\right)%
\end{array}%
}\right) } \\ 
+\exp {\left( {\scriptsize 
\begin{array}{l}
2(s_{u}-1)(s_{u}+s_{v}-1)C_{SNR}\left( M-\overset{M}{\underset{k=1}{\sum }}%
\cos {\left( \frac{2\pi }{\lambda }d_{x_{k}}h_{u}\right) }\right) \\ 
+2(s_{v}-1)(s_{u}+s_{v}-1)C_{SNR}\left( M-\overset{M}{\underset{k=1}{\sum }}%
\cos {\left( \frac{2\pi }{\lambda }d_{y_{k}}h_{v}\right) }\right) \\ 
-2(1-s_{u})(1-s_{v})C_{SNR}\left( M-\overset{M}{\underset{k=1}{\sum }}\cos {%
\left( \frac{2\pi }{\lambda }(d_{x_{k}}h_{u}-d_{y_{k}}h_{v})\right) }\right)%
\end{array}%
}\right) } \\ 
-\exp {\left( {\scriptsize 
\begin{array}{l}
2s_{u}(s_{u}-s_{v})C_{SNR}\left( M-\overset{M}{\underset{k=1}{\sum }}\cos {%
\left( \frac{2\pi }{\lambda }d_{x_{k}}h_{u}\right) }\right) \\ 
+2(1-s_{v})(s_{u}-s_{v})C_{SNR}\left( M-\overset{M}{\underset{k=1}{\sum }}%
\cos {\left( \frac{2\pi }{\lambda }d_{y_{k}}h_{v}\right) }\right) \\ 
+2s_{u}(s_{v}-1)C_{SNR}\left( M-\overset{M}{\underset{k=1}{\sum }}\cos {%
\left( \frac{2\pi }{\lambda }(d_{x_{k}}h_{u}+d_{y_{k}}h_{v})\right) }\right)%
\end{array}%
}\right) } \\ 
-\exp {\left( {\scriptsize 
\begin{array}{l}
2(s_{u}-1)(s_{u}-s_{v})C_{SNR}\left( M-\overset{M}{\underset{k=1}{\sum }}%
\cos {\left( \frac{2\pi }{\lambda }d_{x_{k}}h_{u}\right) }\right) \\ 
+2s_{v}(s_{v}-s_{u})C_{SNR}\left( M-\overset{M}{\underset{k=1}{\sum }}\cos {%
\left( \frac{2\pi }{\lambda }d_{y_{k}}h_{v}\right) }\right) \\ 
+2(s_{u}-1)s_{v}C_{SNR}\left( M-\overset{M}{\underset{k=1}{\sum }}\cos {%
\left( \frac{2\pi }{\lambda }(d_{x_{k}}h_{u}+d_{y_{k}}h_{v})\right) }\right)%
\end{array}%
}\right) }%
\end{array}%
\right) }{\exp {\left( 2s_{u}(s_{u}-1)C_{SNR}\left( M-\overset{M}{\underset{%
k=1}{\sum }}\cos {\left( \frac{2\pi }{\lambda }d_{x_{k}}h_{u}\right) }%
\right) \right) }\exp {\left( 2s_{v}(s_{v}-1)C_{SNR}\left( M-\overset{M}{%
\underset{k=1}{\sum }}\cos {\left( \frac{2\pi }{\lambda }d_{y_{k}}h_{v}%
\right) }\right) \right) }},  \label{eqn: CWWB planar Guv}
\end{equation}%
and $\{\mathbf{G}\}_{uv}=\{\mathbf{G}\}_{vu}$. Consequently, the conditional
Weiss-Weinstein bound is $2\times 2$ matrix given by using the above
equations in Eqn. (\ref{eqn: UWWB planar with s}). As for the unconditional
case, if we set $s_{u}=s_{v}=1/2,$ one obtains the following simplified
expressions

\begin{eqnarray}
\left\{ \mathbf{G}\right\} _{uu} &=&\frac{2\left( 1-\frac{|h_{u}|}{2}\right)
-2(1-|h_{u}|)\exp {\left( -\frac{C_{SNR}}{2}\left( M-\overset{M}{\underset{%
k=1}{\sum }}\cos {\left( \frac{4\pi }{\lambda }d_{x_{k}}h_{u}\right) }%
\right) \right) }}{\left( 1-\frac{|h_{u}|}{2}\right) ^{2}\exp {\left(
-C_{SNR}\left( M-\overset{M}{\underset{k=1}{\sum }}\cos {\left( \frac{2\pi }{%
\lambda }d_{x_{k}}h_{u}\right) }\right) \right) }}, \\
\left\{ \mathbf{G}\right\} _{vv} &=&\frac{2\left( 1-\frac{|h_{v}|}{2}\right)
-2(1-|h_{v}|)\exp {\left( -\frac{C_{SNR}}{2}\left( M-\overset{M}{\underset{%
k=1}{\sum }}\cos {\left( \frac{4\pi }{\lambda }d_{y_{k}}h_{v}\right) }%
\right) \right) }}{\left( 1-\frac{|h_{v}|}{2}\right) ^{2}\exp {\left(
-C_{SNR}\left( M-\overset{M}{\underset{k=1}{\sum }}\cos {\left( \frac{2\pi }{%
\lambda }d_{y_{k}}h_{v}\right) }\right) \right) }}, \\
\left\{ \mathbf{G}\right\} _{uv} &=&\frac{\left( 
\begin{array}{l}
2\exp {\left( -\frac{C_{SNR}}{2}\left( M-\overset{M}{\underset{k=1}{\sum }}%
\cos {\left( \frac{2\pi }{\lambda }(d_{x_{k}}h_{u}-d_{y_{k}}h_{v})\right) }%
\right) \right) } \\ 
-2\exp {\left( -\frac{C_{SNR}}{2}\left( M-\overset{M}{\underset{k=1}{\sum }}%
\cos {\left( \frac{2\pi }{\lambda }(d_{x_{k}}h_{u}+d_{y_{k}}h_{v})\right) }%
\right) \right) }%
\end{array}%
\right) }{\exp {\left( -\frac{C_{SNR}}{2}\left( 2M-\overset{M}{\underset{k=1}%
{\sum }}\cos {\left( \frac{2\pi }{\lambda }d_{x_{k}}h_{u}\right) }-\overset{M%
}{\underset{k=1}{\sum }}\cos {\left( \frac{2\pi }{\lambda }%
d_{y_{k}}h_{v}\right) }\right) \right) }}.
\end{eqnarray}

By using the above expressions in Eqn. (\ref{eqn: UWWB planar with s}) and
after an optimization over the test points, one obtains the Weiss-Weinstein
bound.

\subsection{Source localization with a non-uniform linear array}

We now briefly consider the DOA estimation of a single narrow band source in
the far area by using a non-uniform linear array antenna. Without loss of
generality, let us assume that the linear array antenna lays on the $Ox$
axis of the coordinate system (see Fig.~\ref{fig:Coordinate}), consequently, 
$d_{y_{i}}=0,$ $\forall i$. The sensor positions vector is denoted $\left[
d_{x_{1}}\ldots d_{x_{M}}\right] .$ By letting $\theta =\sin {\varphi ,}$
where $\varphi $ denotes the elevation angle of the source, the steering
vector is then given by 
\begin{equation}
\mathbf{a}(\theta )=\left[ \exp {\left( j\frac{2\pi }{\lambda }%
d_{x_{1}}\theta \right) }\ldots \exp {\left( j\frac{2\pi }{\lambda }%
d_{x_{M}}\theta \right) }\right] ^{T}.
\end{equation}

We assume that the parameter $\theta $ follows a uniform distribution over $%
\left[ -1,1\right] $. As in Section \ref{Sec: Analysis of eta with uniform
prior} and since the parameter of interest is a scalar, matrix $\mathbf{H}$
of the test points becomes a scalar denoted $h_{\theta }$. In the same way,
there is only one element $s_{i}\in \left[ 0,1\right] $ which will be simply
denoted $s$. The closed-form expressions given here are straightforwardly
obtained from the aforementioned results on the planar array about the
element denoted $\left\{ \mathbf{G}\right\} _{uu}.$ We will continue to use
the previously introduced notations $U_{SNR}=\frac{\sigma _{s}^{4}}{\sigma
_{n}^{2}(M\sigma _{s}^{2}+\sigma _{n}^{2})}$ and $C_{SNR}=\frac{1}{\sigma
_{n}^{2}}\overset{T}{\underset{t=1}{\sum }}{\left\Vert s(t)\right\Vert ^{2}.}
$

\subsubsection{Unconditional observation model $\mathcal{M}_{1}$}

The closed-form expression of the unconditional Weiss-Weinstein bound,
denoted $UWWB$, is given by 
\begin{equation}
UWWB=\frac{h_{\theta }^{2}\left( 1-\frac{|h_{\theta }|}{2}\right) ^{2}\left(
1+s(1-s)U_{SNR}\left( M^{2}-\left\Vert \overset{M}{\underset{k=1}{\sum }}%
\exp {\left( -j\frac{2\pi }{\lambda }d_{x_{k}}h_{\theta }\right) }%
\right\Vert ^{2}\right) \right) ^{-2T}}{\left( 
\begin{array}{l}
\left( 1-\frac{|h_{\theta }|}{2}\right) \left( 
\begin{array}{l}
\left( 1+2s(1-2s)U_{SNR}\left( M^{2}-\left\Vert \overset{M}{\underset{k=1}{%
\sum }}\exp {\left( -j\frac{2\pi }{\lambda }d_{x_{k}}h_{\theta }\right) }%
\right\Vert ^{2}\right) \right) ^{-T} \\ 
+\left( 1+2(1-s)(2s-1)U_{SNR}\left( M^{2}-\left\Vert \overset{M}{\underset{%
k=1}{\sum }}\exp {\left( -j\frac{2\pi }{\lambda }d_{x_{k}}h_{\theta }\right) 
}\right\Vert ^{2}\right) \right) ^{-T}%
\end{array}%
\right) \\ 
-2\left( 1-|h_{\theta }|\right) \left( 1+s(1-s)U_{SNR}\left(
M^{2}-\left\Vert \overset{M}{\underset{k=1}{\sum }}\exp {\left( -j\frac{4\pi 
}{\lambda }d_{x_{k}}h_{\theta }\right) }\right\Vert ^{2}\right) \right) ^{-T}%
\end{array}%
\right) }.  \label{eqn: UWWB LA with s}
\end{equation}

In order to find one optimal value of $s$ that maximizes $\mathbf{H}\mathbf{G%
}^{-1}\mathbf{H}^{T}$, $\forall h_{\theta }$ we have considered the
derivative of $\mathbf{H}\mathbf{G}^{-1}\mathbf{H}^{T}$ w.r.t. $s$. The
calculation (not reported here) is straightforward and it is easy to see
that $\left. \frac{\partial \mathbf{H}\mathbf{G}^{-1}\mathbf{H}^{T}}{%
\partial {s}}\right\vert _{s=\frac{1}{2}}=0.$ Consequently, the
Weiss-Weinstein bound has just to be optimized over $h_{\theta }$ and is
simplified leading to%
\begin{equation}
UWWB={\underset{h_{\theta }}{\sup }}\frac{h_{\theta }^{2}\left( 1-\frac{%
|h_{\theta }|}{2}\right) ^{2}\left( 1+\frac{U_{SNR}}{4}\left(
M^{2}-\left\Vert \overset{M}{\underset{k=1}{\sum }}\exp {\left( -j\frac{2\pi 
}{\lambda }d_{x_{k}}h_{\theta }\right) }\right\Vert ^{2}\right) \right)
^{-2T}}{2\left( 1-\frac{|h_{\theta }|}{2}\right) -2\left( 1-|h_{\theta
}|\right) \left( 1+\frac{U_{SNR}}{4}\left( M^{2}-\left\Vert \overset{M}{%
\underset{k=1}{\sum }}\exp {\left( -j\frac{4\pi }{\lambda }%
d_{x_{k}}h_{\theta }\right) }\right\Vert ^{2}\right) \right) ^{-T}}.
\end{equation}

In the classical case of a uniform linear array (\emph{i.e.}, $d_{x_{k}}=d$%
), this expression can be still simplified by noticing that $\overset{M}{%
\underset{k=1}{\sum }}\exp {\left( -j\frac{2\pi }{\lambda }%
d_{x_{k}}h_{\theta }\right) =}M\exp {\left( -j\frac{2\pi d}{\lambda }%
h_{\theta }\right) .}$

\subsubsection{Conditional observation model $\mathcal{M}_{2}$\label{Sec:
Specific application LA Conditional model}}

The closed-form expression of the conditional Weiss-Weinstein bound $CWWB$
is given by 
\begin{equation}
CWWB=\frac{h_{\theta }^{2}\left( 1-\frac{|h_{\theta }|}{2}\right) ^{2}\exp {%
\left( 4s(s-1)C_{SNR}\left( M-\overset{M}{\underset{k=1}{\sum }}\cos {\left( 
\frac{2\pi }{\lambda }d_{x_{k}}h_{\theta }\right) }\right) \right) }}{\left( 
\begin{array}{l}
\left( 1-\frac{|h_{\theta }|}{2}\right) \left( 
\begin{array}{l}
\exp {\left( 4s(2s-1)C_{SNR}\left( M-\overset{M}{\underset{k=1}{\sum }}\cos {%
\left( \frac{2\pi }{\lambda }d_{x_{k}}h_{\theta }\right) }\right) \right) }
\\ 
+\exp {\left( 4(2s-1)(s-1)C_{SNR}\left( M-\overset{M}{\underset{k=1}{\sum }}%
\cos {\left( \frac{2\pi }{\lambda }d_{x_{k}}h_{\theta }\right) }\right)
\right) }%
\end{array}%
\right) \\ 
-2\left( 1-|h_{\theta }|\right) \exp {\left( 2s(s-1)C_{SNR}\left( M-\overset{%
M}{\underset{k=1}{\sum }}\cos {\left( \frac{4\pi }{\lambda }%
d_{x_{k}}h_{\theta }\right) }\right) \right) }%
\end{array}%
\right) }.  \label{eqn: CWWB LA with s}
\end{equation}

Again, it is easy to check that $\left. \frac{\partial \mathbf{H}\mathbf{G}%
^{-1}\mathbf{H}^{T}}{\partial {s}}\right\vert _{s=\frac{1}{2}}=0.$
Consequently, one optimal value of $s$ that maximizes $\mathbf{H}\mathbf{G}%
^{-1}\mathbf{H}^{T}$, $\forall h_{\theta }$ is $s=\frac{1}{2}.$ The
Weiss-Weinstein bound is then simplified as follows 
\begin{equation}
CWWB=\sup_{h_{\theta }}\frac{h_{\theta }^{2}\left( 1-\frac{|h_{\theta }|}{2}%
\right) ^{2}\exp {\left( -C_{SNR}\left( M-\overset{M}{\underset{k=1}{\sum }}%
\cos {\left( \frac{2\pi }{\lambda }d_{x_{k}}h_{\theta }\right) }\right)
\right) }}{2\left( 1-\frac{|h_{\theta }|}{2}\right) -2\left( 1-|h_{\theta
}|\right) \exp {\left( -\frac{1}{2}C_{SNR}\left( M-\overset{M}{\underset{k=1}%
{\sum }}\cos {\left( \frac{4\pi }{\lambda }d_{x_{k}}h_{\theta }\right) }%
\right) \right) }}.
\end{equation}

In the classical case of a uniform linear array (\emph{i.e.}, $d_{x_{k}}=d$%
), this expression can be still simplified by noticing that $\overset{M}{%
\underset{k=1}{\sum }}\cos {\left( \frac{2\pi }{\lambda }d_{x_{k}}h_{\theta
}\right) =M\cos {\left( \frac{2\pi d}{\lambda }h_{\theta }\right) }.}$

\section{Simulation results and analysis\label{Sec: Simulation}}

As an illustration of the previously derived results, we first consider the
scenario proposed in \cite{BEV96a} Fig. 5, \emph{i.e.}, the DOA estimation
under the unconditional model using an uniform circular array consisting of $%
M=16$ sensors with a half-wavelength inter-sensors spacing. The numbers of
snapshots is $T=100$. Since the array is symmetric, the performance
estimation concerning the parameters $u$ and $v$ are the same, this is why
only the performance with respect to the parameters $u$ is given in Fig. \ref%
{ZZBvsWWB}. The Weiss-Weinstein bound is computed using Eqn. (\ref{eqn: CF
WWB unconditional planar uu}), (\ref{eqn: CF WWB unconditional planar vv})
and (\ref{eqn: CF WWB unconditional planar uv}). The Ziv-Zakai bound is
computed using Eqn. (24) in \cite{BEV96a}. The empirical global mean square
error (MSE) of the maximum \emph{a posteriori} (MAP) estimator is obtained
over $2000$ Monte Carlo trials. As in \cite{BEV96a} Fig. (1b), one observes
that both the Weiss-Weinstein bound and the Ziv-Zakai bound are tight w.r.t.
the MSE of the MAP and capture the SNR threshold. Note that, in \cite{BEV96a}
Fig. (1b), the Weiss-Weinstein bound was computed numerically only.

\begin{figure}[h]
\begin{center}
\epsfig{figure=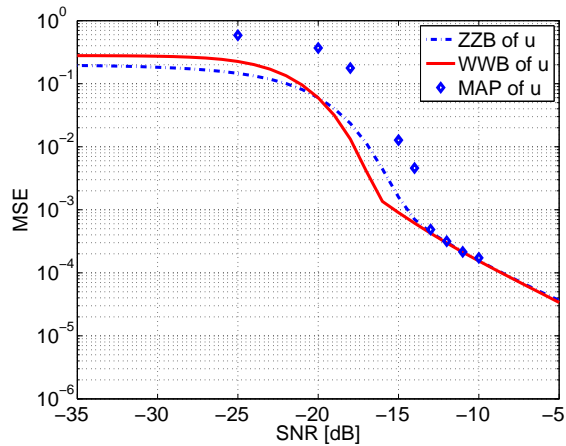,width=0.5\columnwidth}
\end{center}
\caption{Ziv-Zakai bound, Weiss-Weinstein bound and empirical MSE of the MAP
estimator: unconditional case.}
\label{ZZBvsWWB}
\end{figure}

\begin{figure}[h]
\begin{center}
\epsfig{figure=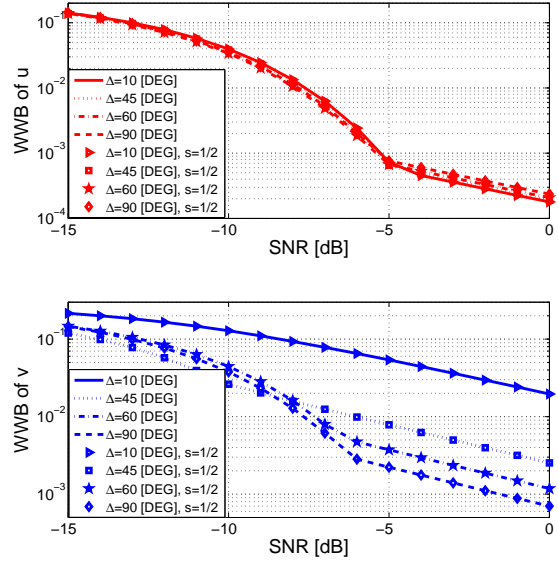,width=0.5\columnwidth}
\end{center}
\caption{Weiss-Weinstein bounds of the V-shaped array w.r.t. the opening
angle $\Delta $.}
\label{V-shaped}
\end{figure}

To the best of our knowledge, their are no closed-form expressions of the
Ziv-Zakai bound for the conditional model available in the literature. In
this case, we consider 3D source localization using a V-shaped array.
Indeed, it has been shown that this kind of array is able to outperform
other classical planar arrays, more particularly the uniform circular array 
\cite{GM06}. This array is made from two branches of uniform linear arrays
with 6 sensors located on each branches and one sensor located at the
origin. We denote $\Delta $ the angle between these two branches. The
sensors are equally spaced with a half-wavelength. The number of snapshots
is $T=20$. Fig. \ref{V-shaped} shows the behavior of the Weiss-Weinstein
bound with respect to the opening angle $\Delta $. One can observe that when 
$\Delta $ varies, the estimation performance concerning the estimation of
parameter $u$ varies slightly. On the contrary, the estimation performance
concerning the estimation of parameter $v$ is strongly dependent on $\Delta $%
. When $\Delta $ increases from $0^{\circ }$ to $90^{\circ }$, the
Weiss-Weinstein bound of $v$ decreases, as well as the SNR threshold. Fig. %
\ref{V-shaped} also shows that $\Delta =90^{\circ }$ is the optimal value,
which is different with the optimal value $\Delta =53.13^{\circ }$ in \cite%
{GM06} since the assumptions concerning the source signal are not the same.

\section{Conclusion\label{Sec: Conclusion}}

In this paper, the Weiss-Weinstein bound on the mean square error has been
studied in the array processing context. In order to analyze the
unconditional and conditional signal source models, the structure of the
bound has been detailed for both Gaussian observation models with
parameterized mean or parameterized covariance matrix.

\appendix

\subsection{Closed-form expression of $\acute{\protect\eta}_{\mathbf{\protect%
\theta }}\left( \protect\alpha ,\protect\beta ,\mathbf{u},\mathbf{v}\right) $
under the Gaussian observation model with parameterized covariance\label%
{Sec: Appendix A}}

Since $\mathbf{y}\left( t\right) \sim \mathcal{CN}\left( \mathbf{0},\mathbf{R%
}_{\mathbf{y}}\left( \mathbf{\theta }\right) \right) $, one has, 
\begin{equation}
\acute{\eta}_{\mathbf{\theta }}\left( \alpha ,\beta ,\mathbf{u},\mathbf{v}%
\right) =\frac{\left\vert \mathbf{R}_{\mathbf{y}}(\mathbf{\theta }%
)\right\vert ^{T(\alpha +\beta -1)}}{\pi ^{MT}\left\vert \mathbf{R}_{\mathbf{%
y}}(\mathbf{\theta +u})\right\vert ^{T\alpha }\left\vert \mathbf{R}_{\mathbf{%
y}}(\mathbf{\theta +v})\right\vert ^{T\beta }}{\underset{\Omega }{\int }}%
\exp {\left( -\overset{T}{\underset{t=1}{\sum }}\mathbf{y}^{H}(t)\mathbf{%
\Gamma }^{-1}\mathbf{y}(t)\right) }d\mathbf{Y},
\end{equation}%
where $\mathbf{\Gamma }^{-1}=\alpha \mathbf{R}_{\mathbf{y}}^{-1}(\mathbf{%
\theta +u})+\beta \mathbf{R}_{\mathbf{y}}^{-1}(\mathbf{\theta +v})-(\alpha
+\beta -1)\mathbf{R}_{\mathbf{y}}^{-1}(\mathbf{\theta }).$ Then, since%
\begin{equation}
{\underset{\Omega }{\int }}\exp {\left\{ -\overset{T}{\underset{t=1}{\sum }}%
\mathbf{y}^{H}(t)\mathbf{\Gamma }^{-1}\mathbf{y}(t)\right\} }d\mathbf{Y}=\pi
^{MT}\left\vert \mathbf{\Gamma }\right\vert ^{T},
\end{equation}%
one has 
\begin{equation}
\acute{\eta}_{\mathbf{\theta }}\left( \alpha ,\beta ,\mathbf{u},\mathbf{v}%
\right) =\frac{\left\vert \mathbf{R}_{\mathbf{y}}(\mathbf{\theta }%
)\right\vert ^{T(\alpha +\beta -1)}\left\vert \mathbf{\Gamma }\right\vert
^{T}}{\left\vert \mathbf{R}_{\mathbf{y}}(\mathbf{\theta +u})\right\vert
^{T\alpha }\left\vert \mathbf{R}_{\mathbf{y}}(\mathbf{\theta +v})\right\vert
^{T\beta }}=\frac{\left\vert \mathbf{R}_{\mathbf{y}}(\mathbf{\theta }%
)\right\vert ^{T(\alpha +\beta -1)}}{\left\vert \mathbf{R}_{\mathbf{y}}(%
\mathbf{\theta +u})\right\vert ^{T\alpha }\left\vert \mathbf{R}_{\mathbf{y}}(%
\mathbf{\theta +v})\right\vert ^{T\beta }\left\vert \mathbf{\Gamma }%
^{-1}\right\vert ^{T}}
\end{equation}

\subsection{Closed-form expression of $\acute{\protect\eta}_{\mathbf{\protect%
\theta }}\left( \protect\alpha ,\protect\beta ,\mathbf{u},\mathbf{v}\right) $
under the Gaussian observation model with parameterized mean\label{Sec:
Appendix B}}

Since $\mathbf{y}\left( t\right) \sim \mathcal{CN}\left( \mathbf{f}%
_{t}\left( \mathbf{\theta }\right) ,\mathbf{R}_{\mathbf{y}}\right) ,$ one
has 
\begin{equation}
\acute{\eta}_{\mathbf{\theta }}\left( \alpha ,\beta ,\mathbf{u},\mathbf{v}%
\right) =\frac{1}{\pi ^{MT}\left\vert \mathbf{R}_{\mathbf{y}}\right\vert ^{T}%
}{\underset{\Omega }{\int }}\exp \left( -\overset{T}{\underset{t=1}{\sum }}%
\xi \left( t\right) \right) d\mathbf{Y},
\end{equation}%
with\footnote{%
For simplicity, the dependance on $t$ of $\mathbf{f}$ and $\mathbf{y}$ is
not emphasized.}%
\begin{eqnarray}
\xi \left( t\right)  &=&\alpha \left( \mathbf{y}-\mathbf{f}_{t}\left( 
\mathbf{\theta +u}\right) \right) ^{H}\mathbf{R}_{\mathbf{y}}^{-1}\left( 
\mathbf{y}-\mathbf{f}_{t}\left( \mathbf{\theta +u}\right) \right) +\beta
\left( \mathbf{y}-\mathbf{f}_{t}\left( \mathbf{\theta +v}\right) \right) ^{H}%
\mathbf{R}_{\mathbf{y}}^{-1}\left( \mathbf{y}-\mathbf{f}_{t}\left( \mathbf{%
\theta +v}\right) \right)   \notag \\
&&+\left( 1-\alpha -\beta \right) \left( \mathbf{y}-\mathbf{f}_{t}\left( 
\mathbf{\theta }\right) \right) ^{H}\mathbf{R}_{\mathbf{y}}^{-1}\left( 
\mathbf{y}-\mathbf{f}_{t}\left( \mathbf{\theta }\right) \right)   \notag \\
&=&\mathbf{y}^{H}\mathbf{R}_{\mathbf{y}}^{-1}\mathbf{y+}\alpha \mathbf{f}%
_{t}^{H}\left( \mathbf{\theta +u}\right) \mathbf{R}_{\mathbf{y}}^{-1}\mathbf{%
f}_{t}\left( \mathbf{\theta +u}\right) \mathbf{+}\beta \mathbf{f}%
_{t}^{H}\left( \mathbf{\theta +v}\right) \mathbf{R}_{\mathbf{y}}^{-1}\mathbf{%
f}_{t}\left( \mathbf{\theta +v}\right) \mathbf{+}\left( 1-\alpha -\beta
\right) \mathbf{f}_{t}^{H}\left( \mathbf{\theta }\right) \mathbf{R}_{\mathbf{%
y}}^{-1}\mathbf{f}_{t}\left( \mathbf{\theta }\right)   \notag \\
&&-2\func{Re}\left\{ \mathbf{y}^{H}\mathbf{R}_{\mathbf{y}}^{-1}\left( \alpha 
\mathbf{f}_{t}\left( \mathbf{\theta +u}\right) +\beta \mathbf{f}_{t}\left( 
\mathbf{\theta +v}\right) +\left( 1-\alpha -\beta \right) \mathbf{f}%
_{t}\left( \mathbf{\theta }\right) \right) \right\} .
\end{eqnarray}

Let us set $\mathbf{x}=\mathbf{y}-\left( \alpha \mathbf{f}_{t}\left( \mathbf{%
\theta +u}\right) +\beta \mathbf{f}_{t}\left( \mathbf{\theta +v}\right)
+\left( 1-\alpha -\beta \right) \mathbf{f}_{t}\left( \mathbf{\theta }\right)
\right) $. Consequently,%
\begin{eqnarray}
\mathbf{x}^{H}\mathbf{R}_{\mathbf{y}}^{-1}\mathbf{x} &=&\mathbf{y}^{H}%
\mathbf{R}_{\mathbf{y}}^{-1}\mathbf{y}-2\func{Re}\left\{ \mathbf{y}^{H}%
\mathbf{R}_{\mathbf{y}}^{-1}\left( \alpha \mathbf{f}_{t}\left( \mathbf{%
\theta +u}\right) +\beta \mathbf{f}_{t}\left( \mathbf{\theta +v}\right)
+\left( 1-\alpha -\beta \right) \mathbf{f}_{t}\left( \mathbf{\theta }\right)
\right) \right\}   \notag \\
&&\hspace{-2cm}\mathbf{+}\left( \alpha \mathbf{f}_{t}^{H}\left( \mathbf{%
\theta +u}\right) +\beta \mathbf{f}_{t}^{H}\left( \mathbf{\theta +v}\right)
+\left( 1-\alpha -\beta \right) \mathbf{f}_{t}^{H}\left( \mathbf{\theta }%
\right) \right) \mathbf{R}_{\mathbf{y}}^{-1}\left( \alpha \mathbf{f}%
_{t}\left( \mathbf{\theta +u}\right) +\beta \mathbf{f}_{t}\left( \mathbf{%
\theta +v}\right) +\left( 1-\alpha -\beta \right) \mathbf{f}_{t}\left( 
\mathbf{\theta }\right) \right) 
\end{eqnarray}

And $\xi \left( t\right) $ can be rewritten as 
\begin{equation}
\xi \left( t\right) =\mathbf{x}^{H}\mathbf{R}_{\mathbf{y}}^{-1}\mathbf{x}+%
\acute{\xi}\left( t\right) ,
\end{equation}%
where%
\begin{eqnarray}
\acute{\xi}\left( t\right)  &=&\alpha \left( 1-\alpha \right) \mathbf{f}%
_{t}^{H}\left( \mathbf{\theta +u}\right) \mathbf{R}_{\mathbf{y}}^{-1}\mathbf{%
f}_{t}\left( \mathbf{\theta +u}\right) \mathbf{+}\beta \left( 1-\beta
\right) \mathbf{f}_{t}^{H}\left( \mathbf{\theta +v}\right) \mathbf{R}_{%
\mathbf{y}}^{-1}\mathbf{f}_{t}\left( \mathbf{\theta +v}\right)   \notag \\
&&\mathbf{+}\left( 1-\alpha -\beta \right) \left( \alpha +\beta \right) 
\mathbf{f}_{t}^{H}\left( \mathbf{\theta }\right) \mathbf{R}_{\mathbf{y}}^{-1}%
\mathbf{f}_{t}\left( \mathbf{\theta }\right) -2\func{Re}\left\{ \alpha \beta 
\mathbf{f}_{t}^{H}\left( \mathbf{\theta +u}\right) \mathbf{R}_{\mathbf{y}%
}^{-1}\mathbf{f}_{t}\left( \mathbf{\theta +v}\right) \right.   \notag \\
&&+\left. \alpha \left( 1-\alpha -\beta \right) \mathbf{f}_{t}^{H}\left( 
\mathbf{\theta +u}\right) \mathbf{R}_{\mathbf{y}}^{-1}\mathbf{f}_{t}\left( 
\mathbf{\theta }\right) +\beta \left( 1-\alpha -\beta \right) \mathbf{f}%
_{t}^{H}\left( \mathbf{\theta +v}\right) \mathbf{R}_{\mathbf{y}}^{-1}\mathbf{%
f}_{t}\left( \mathbf{\theta }\right) \right\} .
\end{eqnarray}

Note that $\acute{\xi}\left( t\right) $ is independent of $\mathbf{x}$. By
defining $\mathbf{X=}\left[ \mathbf{x}\left( 1\right) ,\mathbf{x}\left(
2\right) ,\ldots ,\mathbf{x}\left( T\right) \right] $, the function $\acute{%
\eta}_{\mathbf{\theta }}\left( \alpha ,\beta ,\mathbf{u},\mathbf{v}\right) $
becomes%
\begin{equation}
\acute{\eta}_{\mathbf{\theta }}\left( \alpha ,\beta ,\mathbf{u},\mathbf{v}%
\right) =\frac{1}{\pi ^{MT}\left\vert \mathbf{R}_{\mathbf{y}}\right\vert ^{T}%
}\int_{\Omega }\exp \left( -\overset{T}{\underset{t=1}{\sum }}\mathbf{x}^{H}%
\mathbf{R}_{\mathbf{y}}^{-1}\mathbf{x}+\acute{\xi}\left( t\right) \right) d%
\mathbf{X}=\exp \left( -\overset{T}{\underset{t=1}{\sum }}\acute{\xi}\left(
t\right) \right) ,
\end{equation}%
since $\frac{1}{\pi ^{MT}\left\vert \mathbf{R}_{\mathbf{y}}\right\vert ^{T}}%
\int_{\Omega }\exp \left( -\overset{T}{\underset{t=1}{\sum }}\mathbf{x}^{H}%
\mathbf{R}_{\mathbf{y}}^{-1}\mathbf{x}\right) d\mathbf{X=}1.$

\subsection{Closed-form expressions of $\left\vert {m}_{1}\mathbf{R}_{%
\mathbf{y}}^{-1}(\mathbf{\protect\theta }_{1})+{m}_{2}\mathbf{R}_{\mathbf{y}%
}^{-1}(\mathbf{\protect\theta }_{2})\right\vert $ and $\left\vert {m}_{1}%
\mathbf{R}_{\mathbf{y}}^{-1}(\mathbf{\protect\theta }_{1})+{m}_{2}\mathbf{R}%
_{\mathbf{y}}^{-1}(\mathbf{\protect\theta }_{2})+{m}_{3}\mathbf{R}_{\mathbf{y%
}}^{-1}(\mathbf{\protect\theta }_{3})\right\vert $\label{Sec: Appendix C}}

Note that this calculation is actually an extension of the result obtained
in \cite{XBR04} Appendix A in which ${m}_{1}={m}_{2}=\frac{1}{2}$ and ${m}%
_{3}=0,$ but follows the same method. The inverse of $\mathbf{R}_{\mathbf{y}%
} $ can be deduced from the Woodbury formula 
\begin{equation*}
\mathbf{R}_{\mathbf{y}}^{-1}(\mathbf{\theta })=\frac{1}{\sigma _{n}^{2}}%
\left( \mathbf{I}_{M}-\frac{\sigma _{s}^{2}\mathbf{a}(\mathbf{\theta })%
\mathbf{a}^{H}(\mathbf{\theta })}{\sigma _{s}^{2}\left\Vert \mathbf{a}(%
\mathbf{\theta })\right\Vert ^{2}+\sigma _{n}^{2}}\right) .
\end{equation*}

Then, 
\begin{equation}
\underset{k=1}{\overset{3}{\sum }}{m}_{k}\mathbf{R}_{\mathbf{y}}^{-1}(%
\mathbf{\theta }_{k})=\frac{1}{\sigma _{n}^{2}}\underset{k=1}{\overset{3}{%
\sum }}{m}_{k}\left( \mathbf{I}-\frac{\sigma _{s}^{2}\mathbf{a}(\mathbf{%
\theta }_{k})\mathbf{a}^{H}(\mathbf{\theta }_{k})}{\sigma _{s}^{2}\left\Vert 
\mathbf{a}(\mathbf{\theta }_{k})\right\Vert ^{2}+\sigma _{n}^{2}}\right) .
\label{eqn: CL de covariance inverse}
\end{equation}

Since the rank of $\mathbf{a}(\mathbf{\theta }_{k})\mathbf{a}^{H}(\mathbf{%
\theta }_{k})$ is equal to $1$ and since $\mathbf{\theta }_{1}\neq \mathbf{%
\theta }_{2}\neq \mathbf{\theta }_{3}$ (except for $\mathbf{h}_{k}=\mathbf{h}%
_{l}=\mathbf{0}$), the above matrix has $M-3$ eigenvalues equal to $\frac{1}{%
\sigma _{n}^{2}}\underset{k=1}{\overset{3}{\sum }}{m}_{k}$ and $3$
eigenvalues corresponding to the eigenvectors made from the linear
combination of $\mathbf{a}(\mathbf{\theta }_{1})$, $\mathbf{a}(\mathbf{%
\theta }_{2})$, and $\mathbf{a}(\mathbf{\theta }_{3})$: $\mathbf{a}(\mathbf{%
\theta }_{1})+p\mathbf{a}(\mathbf{\theta }_{2})+q\mathbf{a}(\mathbf{\theta }%
_{3})$. The determinant will then be the product of these $M$ eigenvalues%
\footnote{%
Note that we are only interested by the eigenvalues. Consequently, the
linear combination of of $\mathbf{a}(\mathbf{\theta }_{1})$, $\mathbf{a}(%
\mathbf{\theta }_{2})$, and $\mathbf{a}(\mathbf{\theta }_{3})$ can be
written $\mathbf{a}(\mathbf{\theta }_{1})+p\mathbf{a}(\mathbf{\theta }_{2})+q%
\mathbf{a}(\mathbf{\theta }_{3})$ instead of $r\mathbf{a}(\mathbf{\theta }%
_{1})+p\mathbf{a}(\mathbf{\theta }_{2})+q\mathbf{a}(\mathbf{\theta }_{3})$}.
Let us set 
\begin{equation}
\varphi _{k}=\frac{\sigma _{s}^{2}}{\sigma _{s}^{2}\left\Vert \mathbf{a}(%
\mathbf{\theta }_{k})\right\Vert ^{2}+\sigma _{n}^{2}},\quad k=1,2,3.
\end{equation}

Then, the three aforementioned eigenvalues denoted $\lambda $ must satisfy: 
\begin{equation}
\left( \underset{k=1}{\overset{3}{\sum }}{m}_{k}\mathbf{R}_{\mathbf{y}}^{-1}(%
\mathbf{\theta }_{k})\right) \left( \mathbf{a}(\mathbf{\theta }_{1})+p%
\mathbf{a}(\mathbf{\theta }_{2})+q\mathbf{a}(\mathbf{\theta }_{3})\right)
=\lambda \left( \mathbf{a}(\mathbf{\theta }_{1})+p\mathbf{a}(\mathbf{\theta }%
_{2})+q\mathbf{a}(\mathbf{\theta }_{3})\right) .
\end{equation}

By using Eqn. (\ref{eqn: CL de covariance inverse}) in the above equation
and after a factorization with respect to $\mathbf{a}(\mathbf{\theta }_{1})$%
, $\mathbf{a}(\mathbf{\theta }_{2})$, and $\mathbf{a}(\mathbf{\theta }_{3})$
one obtains

\begin{equation}
\left. 
\begin{array}{l}
\left( x-m_{1}\varphi _{1}\left\Vert \mathbf{a}(\mathbf{\theta }%
_{1})\right\Vert ^{2}-pm_{1}\varphi _{1}\mathbf{a}^{H}(\mathbf{\theta }_{1})%
\mathbf{a}(\mathbf{\theta }_{2})-qm_{1}\varphi _{1}\mathbf{a}^{H}(\mathbf{%
\theta }_{1})\mathbf{a}(\mathbf{\theta }_{3})\right) \mathbf{a}(\mathbf{%
\theta }_{1}) \\ 
+\left( -{m}_{2}\varphi _{2}\mathbf{a}^{H}(\mathbf{\theta }_{2})\mathbf{a}(%
\mathbf{\theta }_{1})+p\left( x-{m}_{2}\varphi _{2}\left\Vert \mathbf{a}(%
\mathbf{\theta }_{2})\right\Vert ^{2}\right) -q{m}_{2}\varphi _{2}\mathbf{a}%
^{H}(\mathbf{\theta }_{2})\mathbf{a}(\mathbf{\theta }_{3})\right) \mathbf{a}(%
\mathbf{\theta }_{2}) \\ 
+\left( -{m}_{3}\varphi _{3}\mathbf{a}^{H}(\mathbf{\theta }_{3})\mathbf{a}(%
\mathbf{\theta }_{1})-{m}_{3}\varphi _{3}p\mathbf{a}^{H}(\mathbf{\theta }%
_{3})\mathbf{a}(\mathbf{\theta }_{2})+q\left( x-{m}_{3}\varphi
_{3}\left\Vert \mathbf{a}(\mathbf{\theta }_{3})\right\Vert ^{2}\right)
\right) \mathbf{a}(\mathbf{\theta }_{3})=0,%
\end{array}%
\right.
\end{equation}%
where\footnote{%
Note that, from Eqn. (\ref{eqn: set of eta prime function unconditional
model}), $\underset{k=1}{\overset{3}{\sum }}{m}_{k}=1.$}%
\begin{equation}
x=1-\sigma _{n}^{2}\lambda .
\end{equation}

Consequently, the coefficients of $\mathbf{a}(\mathbf{\theta }_{1})$, $%
\mathbf{a}(\mathbf{\theta }_{2})$, and $\mathbf{a}(\mathbf{\theta }_{3})$
are equals to zero leading to a system of three equations with two unknown ($%
p$ and $q$). Solving the two first equations to find\footnote{$p$ and $q$
are given by%
\begin{equation}
p=\frac{{m}_{2}\varphi _{2}\mathbf{a}^{H}(\mathbf{\theta }_{2})\left(
m_{1}\varphi _{1}\mathbf{a}(\mathbf{\theta }_{1})\mathbf{a}^{H}(\mathbf{%
\theta }_{1})+\left( x-m_{1}\varphi _{1}\left\Vert \mathbf{a}(\mathbf{\theta 
}_{1})\right\Vert ^{2}\right) \mathbf{I}\right) \mathbf{a}(\mathbf{\theta }%
_{3})}{m_{1}\varphi _{1}\mathbf{a}^{H}(\mathbf{\theta }_{1})\left( {m}%
_{2}\varphi _{2}\mathbf{a}(\mathbf{\theta }_{2})\mathbf{a}^{H}(\mathbf{%
\theta }_{2})+\left( x-{m}_{2}\varphi _{2}\left\Vert \mathbf{a}(\mathbf{%
\theta }_{2})\right\Vert ^{2}\right) \mathbf{I}\right) \mathbf{a}(\mathbf{%
\theta }_{3})},
\end{equation}%
and%
\begin{equation}
q=\frac{\left( x-m_{1}\varphi _{1}\left\Vert \mathbf{a}(\mathbf{\theta }%
_{1})\right\Vert ^{2}\right) \left( x-{m}_{2}\varphi _{2}\left\Vert \mathbf{a%
}(\mathbf{\theta }_{2})\right\Vert ^{2}\right) -m_{1}\varphi _{1}{m}%
_{2}\varphi _{2}\mathbf{a}^{H}(\mathbf{\theta }_{1})\mathbf{a}(\mathbf{%
\theta }_{2})\mathbf{a}^{H}(\mathbf{\theta }_{2})\mathbf{a}(\mathbf{\theta }%
_{1})}{m_{1}\varphi _{1}\mathbf{a}^{H}(\mathbf{\theta }_{1})\left( {m}%
_{2}\varphi _{2}\mathbf{a}(\mathbf{\theta }_{2})\mathbf{a}^{H}(\mathbf{%
\theta }_{2})+\left( x-{m}_{2}\varphi _{2}\left\Vert \mathbf{a}(\mathbf{%
\theta }_{2})\right\Vert ^{2}\right) \mathbf{I}\right) \mathbf{a}(\mathbf{%
\theta }_{3})}.
\end{equation}%
} $p$ and $q$, and applying the solution into the last equation, one obtains
the following polynomial equation of $x$

\begin{eqnarray*}
&&x^{3}-x^{2}\underset{k=1}{\overset{3}{\sum }}{m}_{k}\varphi _{k}\left\Vert 
\mathbf{a}(\mathbf{\theta }_{k})\right\Vert ^{2}-\frac{x}{2}\underset{k=1}{%
\overset{3}{\sum }}\underset{\underset{k^{\prime }\neq k}{k^{\prime }=1}}{%
\overset{3}{\sum }}m_{k}\varphi _{k}m_{k^{\prime }}\varphi _{k^{\prime
}}\left( \left\Vert \mathbf{a}^{H}(\mathbf{\theta }_{k})\mathbf{a}(\mathbf{%
\theta }_{k^{\prime }})\right\Vert ^{2}-\left\Vert \mathbf{a}(\mathbf{\theta 
}_{k})\right\Vert ^{2}\left\Vert \mathbf{a}(\mathbf{\theta }_{k^{\prime
}})\right\Vert ^{2}\right) \\
&&-m_{1}m_{2}m_{3}\varphi _{1}\varphi _{2}\varphi _{3}\left( \left\Vert 
\mathbf{a}(\mathbf{\theta }_{1})\right\Vert ^{2}\left\Vert \mathbf{a}(%
\mathbf{\theta }_{2})\right\Vert ^{2}\left\Vert \mathbf{a}(\mathbf{\theta }%
_{3})\right\Vert ^{2}-\left\Vert \mathbf{a}^{H}(\mathbf{\theta }_{2})\mathbf{%
a}(\mathbf{\theta }_{3})\right\Vert ^{2}\left\Vert \mathbf{a}(\mathbf{\theta 
}_{1})\right\Vert ^{2}\right. \\
&&-\left\Vert \mathbf{a}^{H}(\mathbf{\theta }_{1})\mathbf{a}(\mathbf{\theta }%
_{2})\right\Vert ^{2}\left\Vert \mathbf{a}(\mathbf{\theta }_{3})\right\Vert
^{2}-\left\Vert \mathbf{a}^{H}(\mathbf{\theta }_{3})\mathbf{a}(\mathbf{%
\theta }_{1})\right\Vert ^{2}\left\Vert \mathbf{a}^{H}(\mathbf{\theta }%
_{2})\right\Vert ^{2}+\mathbf{a}^{H}(\mathbf{\theta }_{3})\mathbf{a}(\mathbf{%
\theta }_{2})\mathbf{a}^{H}(\mathbf{\theta }_{1})\mathbf{a}(\mathbf{\theta }%
_{3})\mathbf{a}^{H}(\mathbf{\theta }_{2})\mathbf{a}(\mathbf{\theta }_{1}) \\
&&\left. +\mathbf{a}^{H}(\mathbf{\theta }_{3})\mathbf{a}(\mathbf{\theta }%
_{1})\mathbf{a}^{H}(\mathbf{\theta }_{1})\mathbf{a}(\mathbf{\theta }_{2})%
\mathbf{a}^{H}(\mathbf{\theta }_{2})\mathbf{a}(\mathbf{\theta }_{3})\right)
=0
\end{eqnarray*}

Since we are only interested by the product of the three eigenvalues, we do
not have to solve this polynomial in $\lambda $ and only the opposite of the
last term is required. This leads to Eqn. (\ref{eqn: determinant of
covariance matrix with 2 elements}) with $\overset{3}{\underset{k=1}{\sum }}%
m_{k}=1$. Of course, the closed-form expression of $\left\vert {m}_{1}%
\mathbf{R}_{\mathbf{y}}^{-1}(\mathbf{\theta }_{1})+{m}_{2}\mathbf{R}_{%
\mathbf{y}}^{-1}(\mathbf{\theta }_{2})\right\vert $ is obtained by letting ${%
m}_{3}=0$ and\textbf{\ }$\overset{2}{\underset{k=1}{\sum }}m_{k}=1$ in Eqn. (%
\ref{eqn: determinant of covariance matrix with 3 elements}).

\subsection{Closed-form expressions of $\protect\zeta _{\mathbf{\protect%
\theta }}\left( \mathbf{\protect\mu ,\protect\rho }\right) $\label{Sec:
Appendix D}}

Remind that the function $\zeta _{\mathbf{\theta }}\left( \mathbf{\mu ,\rho }%
\right) $ is defined by Eqn. (\ref{eqn: general function zeta}). Let us
define $p$ as the number of parameters per sources (assumed to be constant
for each sources). Then, without loss of generality, the full parameter
vector $\mathbf{\theta }$ can be decomposed as $\mathbf{\theta }=\left[ 
\mathbf{\theta }_{1}^{T}\ldots \mathbf{\theta }_{N}^{T}\right] ^{T}\ $where $%
\mathbf{\theta }_{i}=\left[ \theta _{i,1}\ldots \theta _{i,p}\right] ^{T},$ $%
i=1,\ldots ,N$ with $q=Np$. Remind that $\mathbf{\mu }=\left[ 0\ldots {\mu
_{i}}\ldots 0\right] ^{T}$ and $\mathbf{\rho }=\left[ 0\ldots {\rho _{j}}%
\ldots 0\right] ^{T}$. It exists two distinct cases to study: when both
index $i$ and $j$ are such that $(m-1)p+1\leq {i}\leq {mp,}$ $m=1,\ldots ,N$
and $(m-1)p+1\leq {j}\leq {mp}$ or when $(m-1)p+1\leq {i}\leq {mp,}$ $%
m=1,\ldots ,N$ and $(n-1)p+1\leq {j}\leq {np,}$ $n=1,\ldots ,N$ with $m\neq {%
n}$. Therefore let us denote: 
\begin{equation}
\left\{ 
\begin{array}{l}
\mathbf{\mu }_{m}=\left[ 0\cdots 0\quad h_{i}\quad 0\cdots 0\right] ^{T}\in 
\mathbb{R}
^{p} \\ 
\mathbf{\rho }_{m}=\left[ 0\cdots 0\quad h_{j}\quad 0\cdots 0\right] ^{T}\in 
\mathbb{R}
^{p}%
\end{array}%
\right. \text{if }\left( m-1\right) p+1\leq i,j\leq mp
\end{equation}%
and 
\begin{equation}
\left\{ 
\begin{array}{l}
\mathbf{\mu }_{m}=\left[ 0\cdots 0\quad h_{i}\quad 0\cdots 0\right] ^{T}\in 
\mathbb{R}
^{p}, \\ 
\mathbf{\rho }_{n}=\left[ 0\cdots 0\quad h_{j}\quad 0\cdots 0\right] ^{T}\in 
\mathbb{R}
^{p},%
\end{array}%
\right. \text{if }\left\{ 
\begin{array}{l}
\left( m-1\right) p+1\leq i\leq mp \\ 
\left( n-1\right) p+1\leq j\leq np%
\end{array}%
,\text{ with }m\neq {n.}\right.
\end{equation}

\subsubsection{The case where $\left( m-1\right) p+1\leq i,j\leq mp$}

In this case, one has: 
\begin{equation}
\mathbf{A}\left( \mathbf{\theta +\mu }\right) -\mathbf{A}\left( \mathbf{%
\theta +\rho }\right) =\left[ \mathbf{0\cdots 0}\quad \mathbf{a}\left( 
\mathbf{\theta }_{m}\mathbf{+\mu }_{m}\right) -\mathbf{a}\left( \mathbf{%
\theta }_{m}\mathbf{+\rho }_{m}\right) \quad \mathbf{0\cdots 0}\right] \in 
\mathbb{C}
^{p\times N},
\end{equation}%
and consequently, 
\begin{equation}
\zeta _{\mathbf{\theta }}\left( \mathbf{\mu ,\rho }\right) =\left\Vert 
\mathbf{R}_{\mathbf{n}}^{-1/2}\left( \mathbf{a}\left( \mathbf{\theta }_{m}%
\mathbf{+\mu }_{m}\right) -\mathbf{a}\left( \mathbf{\theta }_{m}\mathbf{%
+\rho }_{m}\right) \right) \right\Vert ^{2}\overset{T}{\underset{t=1}{\sum }}%
\left\Vert \left\{ \mathbf{s}\left( t\right) \right\} _{m}\right\Vert ^{2}.
\end{equation}

Due to Eqn. (\ref{eqn: general element steering vector}), one has%
\begin{equation*}
\hspace{-9cm}\left\Vert \mathbf{R}_{\mathbf{n}}^{-1/2}\left( \mathbf{a}%
\left( \mathbf{\theta }_{m}\mathbf{+\mu }_{m}\right) -\mathbf{a}\left( 
\mathbf{\theta }_{m}\mathbf{+\rho }_{m}\right) \right) \right\Vert ^{2}=
\end{equation*}%
\begin{eqnarray}
&&\underset{i=1}{\overset{M}{\sum }}\underset{j=1}{\overset{M}{\sum }}%
\left\{ \mathbf{R}_{\mathbf{n}}^{-1}\right\} _{i,j}\exp \left( j\frac{2\pi }{%
\lambda }\left( \mathbf{r}_{j}^{T}-\mathbf{r}_{i}^{T}\right) \mathbf{\theta }%
_{m}\right) \left( \exp \left( -j\frac{2\pi }{\lambda }\mathbf{r}_{i}^{T}%
\mathbf{\mu }_{m}\right) -\exp \left( -j\frac{2\pi }{\lambda }\mathbf{r}%
_{i}^{T}\mathbf{\rho }_{m}\right) \right)  \notag \\
&&\times \left( \exp \left( j\frac{2\pi }{\lambda }\mathbf{r}_{j}^{T}\mathbf{%
\mu }_{m}\right) -\exp \left( j\frac{2\pi }{\lambda }\mathbf{r}_{j}^{T}%
\mathbf{\rho }_{m}\right) \right) .
\end{eqnarray}

In particular, in the case where $\mathbf{R}_{\mathbf{n}}=\sigma _{n}^{2}%
\mathbf{I}$ one obtains 
\begin{equation}
\left\Vert \mathbf{R}_{\mathbf{n}}^{-1/2}\left( \mathbf{a}\left( \mathbf{%
\theta }_{m}\mathbf{+\mu }_{m}\right) -\mathbf{a}\left( \mathbf{\theta }_{m}%
\mathbf{+\rho }_{m}\right) \right) \right\Vert ^{2}=\frac{1}{\sigma _{n}^{2}}%
\underset{i=1}{\overset{M}{\sum }}\left\Vert \exp \left( -j\frac{2\pi }{%
\lambda }\mathbf{r}_{i}^{T}\mathbf{\mu }_{m}\right) -\exp \left( -j\frac{%
2\pi }{\lambda }\mathbf{r}_{i}^{T}\mathbf{\rho }_{m}\right) \right\Vert ^{2}.
\end{equation}

\subsubsection{The case where $\left( m-1\right) p+1\leq i\leq mp$ and where 
$\left( n-1\right) p+1\leq j\leq np$}

Without loss generality, we assume that $n>m$. Then, 
\begin{eqnarray}
\mathbf{A}\left( \mathbf{\theta +\mu }\right) -\mathbf{A}\left( \mathbf{%
\theta +\rho }\right) &=&\left[ \mathbf{a}\left( \mathbf{\theta }_{1}\right)
-\mathbf{a}\left( \mathbf{\theta }_{1}\right) \cdots \mathbf{a}\left( 
\mathbf{\theta }_{m}+\mathbf{\mu }_{m}\right) -\mathbf{a}\left( \mathbf{%
\theta }_{m}\right) \cdots \mathbf{a}\left( \mathbf{\theta }_{n}\right) -%
\mathbf{a}\left( \mathbf{\theta }_{n}+\mathbf{\rho }_{n}\right) \cdots 
\mathbf{a}\left( \mathbf{\theta }_{N}\right) -\mathbf{a}\left( \mathbf{%
\theta }_{N}\right) \right]  \notag \\
&=&\left[ \mathbf{0\cdots 0}\quad \mathbf{a}\left( \mathbf{\theta }_{m}%
\mathbf{+\mu }_{m}\right) -\mathbf{a}\left( \mathbf{\theta }_{m}\right)
\quad \mathbf{0\cdots 0}\quad \mathbf{a}\left( \mathbf{\theta }_{m}\right) -%
\mathbf{a}\left( \mathbf{\theta }_{n}\mathbf{+\rho }_{n}\right) \quad 
\mathbf{0\cdots 0}\right] ,
\end{eqnarray}%
and consequently,%
\begin{equation}
\zeta _{\mathbf{\theta }}\left( \mathbf{\mu ,\rho }\right) =\overset{T}{%
\underset{t=1}{\sum }}\left\Vert \mathbf{R}_{\mathbf{n}}^{-1/2}\left( 
\mathbf{a}\left( \mathbf{\theta }_{m}\mathbf{+\mu }_{m}\right) -\mathbf{a}%
\left( \mathbf{\theta }_{m}\right) \right) \left\{ \mathbf{s}\left( t\right)
\right\} _{m}+\left( \mathbf{a}\left( \mathbf{\theta }_{n}\right) -\mathbf{a}%
\left( \mathbf{\theta }_{n}\mathbf{+\rho }_{n}\right) \right) \left\{ 
\mathbf{s}\left( t\right) \right\} _{n}\right\Vert ^{2}.
\end{equation}

Let us set $\mathbf{\varkappa }=\mathbf{R}_{\mathbf{n}}^{-1/2}\left( \mathbf{%
a}\left( \mathbf{\theta }_{m}\mathbf{+\mu }_{m}\right) -\mathbf{a}\left( 
\mathbf{\theta }_{m}\right) \right) $and $\mathbf{\varrho }=\mathbf{R}_{%
\mathbf{n}}^{-1/2}\left( \mathbf{a}\left( \mathbf{\theta }_{n}\right) -%
\mathbf{a}\left( \mathbf{\theta }_{n}\mathbf{+\rho }_{n}\right) \right) .$
Then, $\zeta _{\mathbf{\theta }}\left( \mathbf{\mu ,\rho }\right) $ can be
rewritten%
\begin{eqnarray}
\zeta _{\mathbf{\theta }}\left( \mathbf{\mu ,\rho }\right) &=&\overset{T}{%
\underset{t=1}{\sum }}\left\Vert \mathbf{\varkappa }\left\{ \mathbf{s}\left(
t\right) \right\} _{m}+\mathbf{\varrho }\left\{ \mathbf{s}\left( t\right)
\right\} _{n}\right\Vert ^{2}  \notag \\
&=&\overset{T}{\underset{t=1}{\sum }}\left( \mathbf{\varkappa }^{H}\mathbf{%
\varkappa }\left\Vert \left\{ \mathbf{s}\left( t\right) \right\}
_{m}\right\Vert ^{2}+\mathbf{\varkappa }^{H}\mathbf{\varrho }\left\{ \mathbf{%
s}\left( t\right) \right\} _{m}^{\ast }\left\{ \mathbf{s}\left( t\right)
\right\} _{n}+\mathbf{\varrho }^{H}\mathbf{\varkappa }\left\{ \mathbf{s}%
\left( t\right) \right\} _{m}\left\{ \mathbf{s}\left( t\right) \right\}
_{n}^{\ast }+\mathbf{\varrho }^{H}\mathbf{\varrho }\left\Vert \left\{ 
\mathbf{s}\left( t\right) \right\} _{n}\right\Vert ^{2}\right)  \notag \\
&=&\mathbf{\varkappa }^{H}\mathbf{\varkappa }\overset{T}{\underset{t=1}{\sum 
}}\left\Vert \left\{ \mathbf{s}\left( t\right) \right\} _{m}\right\Vert ^{2}+%
\mathbf{\varrho }^{H}\mathbf{\varrho }\overset{T}{\underset{t=1}{\sum }}%
\left\Vert \left\{ \mathbf{s}\left( t\right) \right\} _{n}\right\Vert ^{2}+2%
\func{Re}\left( \mathbf{\varkappa }^{H}\mathbf{\varrho }\overset{T}{\underset%
{t=1}{\sum }}\left\{ \mathbf{s}\left( t\right) \right\} _{m}^{\ast }\left\{ 
\mathbf{s}\left( t\right) \right\} _{n}\right) .
\end{eqnarray}

By using the structure of the steering matrix $\mathbf{A}$, it leads to 
\begin{equation}
\left\{ 
\begin{array}{l}
\mathbf{\varkappa }^{H}\mathbf{\varkappa }=\underset{i=1}{\overset{M}{\sum }}%
\underset{j=1}{\overset{M}{\sum }}\left\{ \mathbf{R}_{\mathbf{n}%
}^{-1}\right\} _{i,j}\exp \left( j\frac{2\pi }{\lambda }\left( \mathbf{r}%
_{j}^{T}-\mathbf{r}_{i}^{T}\right) \mathbf{\theta }_{m}\right) \exp \left( -j%
\frac{2\pi }{\lambda }\mathbf{r}_{i}^{T}\mathbf{\mu }_{m}\right) \exp \left(
j\frac{2\pi }{\lambda }\mathbf{r}_{j}^{T}\mathbf{\mu }_{m}\right) , \\ 
\mathbf{\varrho }^{H}\mathbf{\varrho }=\underset{i=1}{\overset{M}{\sum }}%
\underset{j=1}{\overset{M}{\sum }}\left\{ \mathbf{R}_{\mathbf{n}%
}^{-1}\right\} _{i,j}\exp \left( j\frac{2\pi }{\lambda }\left( \mathbf{r}%
_{j}^{T}-\mathbf{r}_{i}^{T}\right) \mathbf{\theta }_{n}\right) \exp \left( -j%
\frac{2\pi }{\lambda }\mathbf{r}_{i}^{T}\mathbf{\rho }_{n}\right) \exp
\left( j\frac{2\pi }{\lambda }\mathbf{r}_{j}^{T}\mathbf{\rho }_{n}\right) ,
\\ 
\mathbf{\varkappa }^{H}\mathbf{\varrho }=-\underset{i=1}{\overset{M}{\sum }}%
\underset{j=1}{\overset{M}{\sum }}\left\{ \mathbf{R}_{\mathbf{n}%
}^{-1}\right\} _{i,j}\exp \left( j\frac{2\pi }{\lambda }\left( \mathbf{r}%
_{j}^{T}\mathbf{\theta }_{n}-\mathbf{r}_{i}^{T}\mathbf{\theta }_{m}\right)
\right) \exp \left( -j\frac{2\pi }{\lambda }\mathbf{r}_{i}^{T}\mathbf{\mu }%
_{m}\right) \exp \left( j\frac{2\pi }{\lambda }\mathbf{r}_{j}^{T}\mathbf{%
\rho }_{n}\right) .%
\end{array}%
\right.
\end{equation}

\subsection{Proof of Eqn. (\protect\ref{eqn: UWWB planar Guu}), (\protect\ref%
{eqn: UWWB planar Gvv}) and (\protect\ref{eqn: UWWB planar Guv})\label{Sec:
Appendix E}}

In fact, one only has to prove Eqn. (\ref{eqn: UWWB planar Guv}) since Eqn. (%
\ref{eqn: UWWB planar Guu}) and (\ref{eqn: UWWB planar Gvv}) can be obtained
by letting $h_{u}=h_{v}$ and $s_{u}=s_{v}$ in Eqn. (\ref{eqn: UWWB planar
Guv}) and by using $\left( h_{u},s_{u}\right) $ for Eqn. (\ref{eqn: UWWB
planar Guu}) and $\left( h_{v},s_{v}\right) $ for Eqn. (\ref{eqn: UWWB
planar Gvv}). By plugging Eqn. (\ref{eqn: determinant of covariance matrix
with 1 elements}) and (\ref{eqn: determinant of covariance matrix with 3
elements}) into Eqn. (\ref{eqn: set of eta prime function unconditional
model}), and by considering the following expressions 
\begin{equation*}
\begin{array}{l}
\mathbf{a}^{H}(\mathbf{\theta }+\mathbf{h}_{u})\mathbf{a}(\mathbf{\theta }+%
\mathbf{h}_{v})=\overset{M}{\underset{i=1}{\sum }}\exp {\left( {j}\frac{2\pi 
}{\lambda }(d_{y_{i}}h_{v}-d_{x_{i}}h_{u})\right) =}\left( \mathbf{a}^{H}(%
\mathbf{\theta }+\mathbf{h}_{v})\mathbf{a}(\mathbf{\theta }+\mathbf{h}%
_{u})\right) ^{H}, \\ 
\mathbf{a}^{H}(\mathbf{\theta }\pm \mathbf{h}_{u})\mathbf{a}(\mathbf{\theta }%
)=\overset{M}{\underset{i=1}{\sum }}\exp {\left( \mp {j}\frac{2\pi }{\lambda 
}d_{x_{i}}h_{u}\right) },\text{ and }\mathbf{a}^{H}(\mathbf{\theta }+\mathbf{%
h}_{u})\mathbf{a}(\mathbf{\theta }-\mathbf{h}_{u})=\overset{M}{\underset{i=1}%
{\sum }}\exp {\left( -{j}\frac{4\pi }{\lambda }d_{x_{i}}h_{u}\right) },%
\end{array}%
\end{equation*}%
one obtains the closed-form expressions for the set of functions $\acute{\eta%
}_{\mathbf{\theta }}\left( \alpha ,\beta ,\mathbf{u},\mathbf{v}\right) $

\begin{equation}
\acute{\eta}_{\mathbf{\theta }}(s_{u},s_{v},\mathbf{h}_{u},\mathbf{h}%
_{v})=\left( 
\begin{array}{l}
1-U_{SNR}\left( 
\begin{array}{l}
s_{u}s_{v}\left( \left\Vert \overset{M}{\underset{k=1}{\sum }}\exp {\left( -j%
\frac{2\pi }{\lambda }(d_{x_{k}}h_{u}-d_{y_{k}}h_{v})\right) }\right\Vert
^{2}-M^{2}\right) \\ 
+s_{u}(1-s_{u}-s_{v})\left( \left\Vert \overset{M}{\underset{k=1}{\sum }}%
\exp {\left( -j\frac{2\pi }{\lambda }d_{x_{k}}h_{u}\right) }\right\Vert
^{2}-M^{2}\right) \\ 
+s_{v}(1-s_{u}-s_{v})\left( \left\Vert \overset{M}{\underset{k=1}{\sum }}%
\exp {\left( -j\frac{2\pi }{\lambda }d_{y_{k}}h_{v}\right) }\right\Vert
^{2}-M^{2}\right)%
\end{array}%
\right) \\ 
-s_{u}s_{v}(1-s_{u}-s_{v})\frac{U_{SNR}^{2}\sigma _{n}^{2}}{\sigma _{s}^{2}}%
\times \\ 
\times \left( 
\begin{array}{l}
\overset{M}{\underset{k=1}{\sum }}\exp {\left( j\frac{2{\pi }d_{y_{k}}h_{v}}{%
\lambda }\right) }\overset{M}{\underset{k=1}{\sum }}\exp {\left( -j\frac{2{%
\pi }d_{x_{k}}h_{u}}{\lambda }\right) }\overset{M}{\underset{k=1}{\sum }}%
\exp {\left( j\frac{2\pi (d_{x_{k}}h_{u}-d_{y_{k}}h_{v})}{\lambda }\right) }
\\ 
+\overset{M}{\underset{k=1}{\sum }}\exp {\left( -j\frac{2{\pi }d_{y_{k}}h_{v}%
}{\lambda }\right) }\overset{M}{\underset{k=1}{\sum }}\exp {\left( j\frac{2{%
\pi }d_{x_{k}}h_{u}}{\lambda }\right) }\overset{M}{\underset{k=1}{\sum }}%
\exp {\left( -j\frac{2\pi (d_{x_{k}}h_{u}-d_{y_{k}}h_{v})}{\lambda }\right) }
\\ 
-M\left\Vert \overset{M}{\underset{k=1}{\sum }}\exp {\left( -j\frac{2\pi }{%
\lambda }d_{y_{k}}h_{v}\right) }\right\Vert ^{2}-M\left\Vert \overset{M}{%
\underset{k=1}{\sum }}\exp {\left( -j\frac{2\pi }{\lambda }%
d_{x_{k}}h_{u}\right) }\right\Vert ^{2} \\ 
-M\left\Vert \overset{M}{\underset{k=1}{\sum }}\exp {\left( -j\frac{2\pi }{%
\lambda }(d_{x_{k}}h_{u}-d_{y_{k}}h_{v})\right) }\right\Vert ^{2}+M^{3}%
\end{array}%
\right)%
\end{array}%
\right) ^{-T}
\end{equation}%
\begin{equation}
\begin{array}{l}
\acute{\eta}_{\mathbf{\theta }}(1-s_{u},1-s_{v},-\mathbf{h}_{u},-\mathbf{h}%
_{v})= \\ 
\qquad \left( 
\begin{array}{l}
1-U_{SNR}\left( 
\begin{array}{l}
(1-s_{u})(1-s_{v})\left( \left\Vert \overset{M}{\underset{k=1}{\sum }}\exp {%
\left( j\frac{2\pi }{\lambda }(d_{x_{k}}h_{u}-d_{y_{k}}h_{v})\right) }%
\right\Vert ^{2}-M^{2}\right) \\ 
+(1-s_{u})(s_{u}+s_{v}-1)\left( \left\Vert \overset{M}{\underset{k=1}{\sum }}%
\exp {\left( j\frac{2\pi }{\lambda }d_{x_{k}}h_{u}\right) }\right\Vert
^{2}-M^{2}\right) \\ 
+(1-s_{v})(s_{u}+s_{v}-1)\left( \left\Vert \overset{M}{\underset{k=1}{\sum }}%
\exp {\left( j\frac{2\pi }{\lambda }d_{y_{k}}h_{v}\right) }\right\Vert
^{2}-M^{2}\right)%
\end{array}%
\right) \\ 
-(1-s_{u})(1-s_{v})(s_{u}+s_{v}-1)\frac{U_{SNR}^{2}\sigma _{n}^{2}}{\sigma
_{s}^{2}}\times \\ 
\times \left( 
\begin{array}{l}
\overset{M}{\underset{k=1}{\sum }}\exp {\left( j\frac{2{\pi }d_{y_{k}}h_{v}}{%
\lambda }\right) }\overset{M}{\underset{k=1}{\sum }}\exp {\left( -j\frac{2{%
\pi }d_{x_{k}}h_{u}}{\lambda }\right) }\overset{M}{\underset{k=1}{\sum }}%
\exp {\left( j\frac{2\pi (d_{x_{k}}h_{u}-d_{y_{k}}h_{v})}{\lambda }\right) }
\\ 
+\overset{M}{\underset{k=1}{\sum }}\exp {\left( -j\frac{2{\pi }d_{y_{k}}h_{v}%
}{\lambda }\right) }\overset{M}{\underset{k=1}{\sum }}\exp {\left( j\frac{2{%
\pi }d_{x_{k}}h_{u}}{\lambda }\right) }\overset{M}{\underset{k=1}{\sum }}%
\exp {\left( -j\frac{2\pi (d_{x_{k}}h_{u}-d_{y_{k}}h_{v})}{\lambda }\right) }
\\ 
-M\left\Vert \overset{M}{\underset{k=1}{\sum }}\exp {\left( -j\frac{2\pi }{%
\lambda }d_{y_{k}}h_{v}\right) }\right\Vert ^{2}-M\left\Vert \overset{M}{%
\underset{k=1}{\sum }}\exp {\left( -j\frac{2\pi }{\lambda }%
d_{x_{k}}h_{u}\right) }\right\Vert ^{2} \\ 
-M\left\Vert \overset{M}{\underset{k=1}{\sum }}\exp {\left( -j\frac{2\pi }{%
\lambda }(d_{x_{k}}h_{u}-d_{y_{k}}h_{v})\right) }\right\Vert ^{2}+M^{3}%
\end{array}%
\right)%
\end{array}%
\right) ^{-T}%
\end{array}%
\end{equation}%
\begin{equation}
\begin{array}{l}
\acute{\eta}_{\mathbf{\theta }}(s_{u},1-s_{v},\mathbf{h}_{u},-\mathbf{h}%
_{v})= \\ 
\qquad \left( 
\begin{array}{l}
1-U_{SNR}\left( 
\begin{array}{l}
s_{u}(1-s_{v})\left( \left\Vert \overset{M}{\underset{k=1}{\sum }}\exp {%
\left( -j\frac{2\pi }{\lambda }(d_{x_{k}}h_{u}+d_{y_{k}}h_{v})\right) }%
\right\Vert ^{2}-M^{2}\right) \\ 
+s_{u}(s_{v}-s_{u})\left( \left\Vert \overset{M}{\underset{k=1}{\sum }}\exp {%
\left( -j\frac{2\pi }{\lambda }d_{x_{k}}h_{u}\right) }\right\Vert
^{2}-M^{2}\right) \\ 
+(1-s_{v})(s_{v}-s_{u})\left( \left\Vert \overset{M}{\underset{k=1}{\sum }}%
\exp {\left( j\frac{2\pi }{\lambda }d_{y_{k}}h_{v}\right) }\right\Vert
^{2}-M^{2}\right)%
\end{array}%
\right) \\ 
-s_{u}(1-s_{v})(s_{v}-s_{u})\frac{U_{SNR}^{2}\sigma _{n}^{2}}{\sigma _{s}^{2}%
}\times \\ 
\times \left( 
\begin{array}{l}
\overset{M}{\underset{k=1}{\sum }}\exp {\left( j\frac{2{\pi }d_{y_{k}}h_{v}}{%
\lambda }\right) }\overset{M}{\underset{k=1}{\sum }}\exp {\left( j\frac{2{%
\pi }d_{x_{k}}h_{u}}{\lambda }\right) }\overset{M}{\underset{k=1}{\sum }}%
\exp {\left( -j\frac{2\pi (d_{x_{k}}h_{u}+d_{y_{k}}h_{v})}{\lambda }\right) }
\\ 
+\overset{M}{\underset{k=1}{\sum }}\exp {\left( -j\frac{2{\pi }d_{y_{k}}h_{v}%
}{\lambda }\right) }\overset{M}{\underset{k=1}{\sum }}\exp {\left( -j\frac{2{%
\pi }d_{x_{k}}h_{u}}{\lambda }\right) }\overset{M}{\underset{k=1}{\sum }}%
\exp {\left( j\frac{2\pi (d_{x_{k}}h_{u}+d_{y_{k}}h_{v})}{\lambda }\right) }
\\ 
-M\left\Vert \overset{M}{\underset{k=1}{\sum }}\exp {\left( j\frac{2\pi }{%
\lambda }d_{y_{k}}h_{v}\right) }\right\Vert ^{2}-M\left\Vert \overset{M}{%
\underset{k=1}{\sum }}\exp {\left( -j\frac{2\pi }{\lambda }%
d_{x_{k}}h_{u}\right) }\right\Vert ^{2} \\ 
-M\left\Vert \overset{M}{\underset{k=1}{\sum }}\exp {\left( -j\frac{2\pi }{%
\lambda }(d_{x_{k}}h_{u}+d_{y_{k}}h_{v})\right) }\right\Vert ^{2}+M^{3}%
\end{array}%
\right)%
\end{array}%
\right) ^{-T}%
\end{array}%
\end{equation}%
\begin{equation}
\begin{array}{l}
\acute{\eta}_{\mathbf{\theta }}(1-s_{u},s_{v},-\mathbf{h}_{u},\mathbf{h}%
_{v})= \\ 
\qquad \left( 
\begin{array}{l}
1-U_{SNR}\left( 
\begin{array}{l}
s_{v}(1-s_{u})\left( \left\Vert \overset{M}{\underset{k=1}{\sum }}\exp {%
\left( -j\frac{2\pi }{\lambda }(d_{x_{k}}h_{u}+d_{y_{k}}h_{v})\right) }%
\right\Vert ^{2}-M^{2}\right) \\ 
+s_{v}(s_{u}-s_{v})\left( \left\Vert \overset{M}{\underset{k=1}{\sum }}\exp {%
\left( -j\frac{2\pi }{\lambda }d_{x_{k}}h_{u}\right) }\right\Vert
^{2}-M^{2}\right) \\ 
+(1-s_{u})(s_{u}-s_{v})\left( \left\Vert \overset{M}{\underset{k=1}{\sum }}%
\exp {\left( -j\frac{2\pi }{\lambda }d_{y_{k}}h_{v}\right) }\right\Vert
^{2}-M^{2}\right)%
\end{array}%
\right) \\ 
-s_{v}(1-s_{u})(s_{u}-s_{v})\frac{U_{SNR}^{2}\sigma _{n}^{2}}{\sigma _{s}^{2}%
}\times \\ 
\times \left( 
\begin{array}{l}
\overset{M}{\underset{k=1}{\sum }}\exp {\left( j\frac{2{\pi }d_{y_{k}}h_{v}}{%
\lambda }\right) }\overset{M}{\underset{k=1}{\sum }}\exp {\left( j\frac{2{%
\pi }d_{x_{k}}h_{u}}{\lambda }\right) }\overset{M}{\underset{k=1}{\sum }}%
\exp {\left( -j\frac{2\pi (d_{x_{k}}h_{u}+d_{y_{k}}h_{v})}{\lambda }\right) }
\\ 
+\overset{M}{\underset{k=1}{\sum }}\exp {\left( -j\frac{2{\pi }d_{y_{k}}h_{v}%
}{\lambda }\right) }\overset{M}{\underset{k=1}{\sum }}\exp {\left( -j\frac{2{%
\pi }d_{x_{k}}h_{u}}{\lambda }\right) }\overset{M}{\underset{k=1}{\sum }}%
\exp {\left( j\frac{2\pi (d_{x_{k}}h_{u}+d_{y_{k}}h_{v})}{\lambda }\right) }
\\ 
-M\left\Vert \overset{M}{\underset{k=1}{\sum }}\exp {\left( -j\frac{2\pi }{%
\lambda }d_{y_{k}}h_{v}\right) }\right\Vert ^{2}-M\left\Vert \overset{M}{%
\underset{k=1}{\sum }}\exp {\left( -j\frac{2\pi }{\lambda }%
d_{x_{k}}h_{u}\right) }\right\Vert ^{2} \\ 
-M\left\Vert \overset{M}{\underset{k=1}{\sum }}\exp {\left( -j\frac{2\pi }{%
\lambda }(d_{x_{k}}h_{u}+d_{y_{k}}h_{v})\right) }\right\Vert ^{2}+M^{3}%
\end{array}%
\right)%
\end{array}%
\right) ^{-T}%
\end{array}%
\end{equation}%
\begin{equation}
\acute{\eta}_{\mathbf{\theta }}(s_{u},0,\mathbf{h}_{u},\mathbf{0})=\left(
1+s_{u}(1-s_{u})U_{SNR}\left( M^{2}-\left\Vert \overset{M}{\underset{k=1}{%
\sum }}\exp {\left( -j\frac{2\pi }{\lambda }d_{x_{k}}h_{u}\right) }%
\right\Vert ^{2}\right) \right) ^{-T},
\end{equation}%
and%
\begin{equation}
\acute{\eta}_{\mathbf{\theta }}(0,s_{v},\mathbf{0},\mathbf{h}_{v})=\left(
1+s_{v}(1-s_{v})U_{SNR}\left( M^{2}-\left\Vert \overset{M}{\underset{k=1}{%
\sum }}\exp {\left( -j\frac{2\pi }{\lambda }d_{y_{k}}h_{v}\right) }%
\right\Vert ^{2}\right) \right) ^{-T}.
\end{equation}

One notices that the set of functions $\acute{\eta}_{\mathbf{\theta }}\left(
\alpha ,\beta ,\mathbf{u},\mathbf{v}\right) $ does not depend on $\mathbf{%
\theta }$. Consequently, it is also easy to obtain the Weiss-Weinstein bound
(throughout the set of functions $\eta \left( \alpha ,\beta ,\mathbf{u},%
\mathbf{v}\right) $) by using the results of Section \ref{Sec: Analysis of
eta with uniform prior} whatever the considered prior on $\theta $ (only the
integral $\int_{\Theta }\frac{p^{\alpha +\beta }\left( \theta +u\right) }{%
p^{\alpha +\beta -1}\left( \theta \right) }d\theta $ has to be calculated or
computed numerically). In our case of a uniform prior, the results are
straightforward and leads to Eqn. (\ref{eqn: UWWB planar Guu}), (\ref{eqn:
UWWB planar Gvv}) and (\ref{eqn: UWWB planar Guv}).

\subsection{Proof of Eqn. (\protect\ref{eqn: CWWB planar Guu}), (\protect\ref%
{eqn: CWWB planar Gvv}) and (\protect\ref{eqn: CWWB planar Guv})\label{Sec:
Appendix F}}

The set of functions $\acute{\eta}_{\mathbf{\theta }}\left( \alpha ,\beta ,%
\mathbf{u},\mathbf{v}\right) $ is given by Eqn. (\ref{eqn: set of eta prime
function conditional model}). So, it only remains the calculation of
functions $\zeta _{\mathbf{\theta }}\left( \mathbf{\mu ,\rho }\right) $ from
Eqn. (\ref{eqn: general function zeta}). Since $\mathbf{R}_{\mathbf{n}%
}=\sigma _{n}^{2}\mathbf{I,}$ one obtains

\begin{equation}
\left\{ 
\begin{array}{lllll}
\zeta _{\mathbf{\theta }}(\mathbf{h}_{u},\mathbf{0}) & = & \zeta _{\mathbf{%
\theta }}(-\mathbf{h}_{u},\mathbf{0}) & = & 2C_{SNR}\left( M-\overset{M}{%
\underset{k=1}{\sum }}\cos {\left( \frac{2\pi }{\lambda }d_{xk}h_{u}\right) }%
\right) , \\ 
\zeta _{\mathbf{\theta }}(\mathbf{h}_{v},\mathbf{0}) & = & \zeta _{\mathbf{%
\theta }}(-\mathbf{h}_{v},\mathbf{0}) & = & 2C_{SNR}\left( M-\overset{M}{%
\underset{k=1}{\sum }}\cos {\left( \frac{2\pi }{\lambda }d_{yk}h_{v}\right) }%
\right) , \\ 
\zeta _{\mathbf{\theta }}(\mathbf{h}_{u},-\mathbf{h}_{u}) & = & \zeta _{%
\mathbf{\theta }}(-\mathbf{h}_{u},\mathbf{h}_{u}) & = & 2C_{SNR}\left( M-%
\overset{M}{\underset{k=1}{\sum }}\cos {\left( \frac{4\pi }{\lambda }%
d_{xk}h_{u}\right) }\right) , \\ 
\zeta _{\mathbf{\theta }}(\mathbf{h}_{v},-\mathbf{h}_{v}) & = & \zeta _{%
\mathbf{\theta }}(-\mathbf{h}_{v},\mathbf{h}_{v}) & = & 2C_{SNR}\left( M-%
\overset{M}{\underset{k=1}{\sum }}\cos {\left( \frac{4\pi }{\lambda }%
d_{yk}h_{v}\right) }\right) , \\ 
\zeta _{\mathbf{\theta }}(\mathbf{h}_{u},\mathbf{h}_{v}) & = & \zeta _{%
\mathbf{\theta }}(\mathbf{h}_{v},\mathbf{h}_{u}) & = & \zeta _{\mathbf{%
\theta }}(-\mathbf{h}_{u},-\mathbf{h}_{v}) \\ 
& = & \zeta _{\mathbf{\theta }}(-\mathbf{h}_{v},-\mathbf{h}_{u}) & = & 
2C_{SNR}\left( M-\overset{M}{\underset{k=1}{\sum }}\cos {\left( \frac{2\pi }{%
\lambda }(d_{xk}h_{u}-d_{yk}h_{v})\right) }\right) , \\ 
\zeta _{\mathbf{\theta }}(-\mathbf{h}_{u},\mathbf{h}_{v}) & = & \zeta _{%
\mathbf{\theta }}(\mathbf{h}_{u},-\mathbf{h}_{v}) & = & \zeta _{\mathbf{%
\theta }}(\mathbf{h}_{v},-\mathbf{h}_{u}) \\ 
& = & \zeta _{\mathbf{\theta }}(-\mathbf{h}_{v},\mathbf{h}_{u}) & = & 
2C_{SNR}\left( M-\overset{M}{\underset{k=1}{\sum }}\cos {\left( \frac{2\pi }{%
\lambda }(d_{xk}h_{u}+d_{yk}h_{v})\right) }\right) , \\ 
\zeta _{\mathbf{\theta }}(\mathbf{h}_{u},\mathbf{h}_{u}) & = & \zeta _{%
\mathbf{\theta }}(\mathbf{h}_{v},\mathbf{h}_{v}) & = & \zeta _{\mathbf{%
\theta }}(-\mathbf{h}_{u},-\mathbf{h}_{u})=\zeta _{\mathbf{\theta }}(-%
\mathbf{h}_{v},-\mathbf{h}_{v})=0.%
\end{array}%
\right.
\end{equation}

Again, since the set of functions $\zeta _{\mathbf{\theta }}\left( \mathbf{%
\mu ,\rho }\right) $ does not depend on $\mathbf{\theta ,}$ the set of
functions $\acute{\eta}_{\mathbf{\theta }}\left( \alpha ,\beta ,\mathbf{u},%
\mathbf{v}\right) $ is given by plugging the above equations into Eqn. (\ref%
{eqn: set of eta prime function conditional model}) and does not depend on $%
\mathbf{\theta .}$ Consequently, as in unconditional case, the set of
functions $\eta \left( \alpha ,\beta ,\mathbf{u},\mathbf{v}\right) $ is
obtained by using the results of Section \ref{Sec: Analysis of eta with
uniform prior} whatever the considered prior on $\theta $. In our case of a
uniform prior, the results are straightforward and leads to Eqn. (\ref{eqn:
CWWB planar Guu}), (\ref{eqn: CWWB planar Gvv}) and (\ref{eqn: CWWB planar
Guv}).

\bibliographystyle{IEEEtran}
\bibliography{Alex}

\end{document}